\newtheorem{remark}{Remark}

\newtheorem{theorem}{Theorem}
\newtheorem{lemma}{Lemma}

\documentclass[10pt,journal,compsoc]{IEEEtran}
\newcommand{\bea}{\begin{eqnarray}}
\newcommand{\eea}{\end{eqnarray}}

\usepackage[dvips]{graphicx}

\usepackage{amsmath,amssymb,amsfonts}
\usepackage{cite}
\usepackage{graphicx}
\usepackage{subfigure}
\usepackage{color}
\usepackage{geometry}
\begin{document}

\title{Complex-order Derivative and Integral Filters and its Applications}

\author{Yiguang Liu
\IEEEcompsocitemizethanks{\IEEEcompsocthanksitem Yiguang Liu is with Vision and Image Processing Laboratory, College of Computer Science, Sichuan University, Chengdu, Sichuan
Province, China, 610065. E-mail: liuyg@scu.edu.cn
}}
\maketitle
\begin{abstract}
In this paper, complex-order derivative and integral filters are proposed, which are consistent with the filters with fractional derivative and integral orders. Compared with the filters designed only with real orders, complex order filters can reveal more details of input signals, and this can benefit a large number of tasks. The tremendous effect of the proposed complex-order filters, has been verified by several image processing examples. Especially, for the challenging problem indicating the diseased regions in prostate TRUS images, the proposed complex order filters look very promising. It is very astonishing that, indicating the diseased regions is fulfilled by complex-order integral, not by derivative. This is completely against the traditional views that using derivative operators to achieve the goal.  
\end{abstract}


\section{Introduction}
\IEEEPARstart{F}{ilters} aims at extracting components or features from a signal, and is widely used in image processing\cite{1030}, computer vision, bio-informatics\cite{1034},
financial modeling\cite{850}, control systems\cite{1029}, seismic data processing\cite{1033}, etc. Filters are so important and fundamental, that there are too many to have a
simple hierarchical classification. Filters based on traditional integer-order calculus demonstrate shortcomings such as ringing artifacts and staircase effects \cite{1030}.
Thus many present-day known filters are resulted from fractional derivatives or integrals, usually called fractional calculus, as a generalization of the traditional integer
calculus and leading to a much wider applicability \cite{858}. Fractional calculus can date back to Cauchy, Riemann, Liouville and Letnikov in 19th century,  and is related to
the letters between Leibniz and Bernoulli about the meaning of derivative of order 1/2 of the power function. Comprehensive reviews on fractional derivatives can refer to
\cite{1041}\cite{1043}.

Indeed, fractional calculus is local for integer order, and non-local for non-integer order\cite{1043}. Fractional calculus is a powerful tool for modeling various phenomena in
mechanics, physics, biology, chemistry, medicine, economy, etc. Fractional calculus stands out in modelling problems due to the concepts of non-locality and memory effect, which
are not well explained by the integer-order calculus. Fractional calculus becomes a source of not only scientific discussion and progress, but also some controversy under the
light of recent proposals\cite{1043}. Actually, in the last few decades, a rapid expansion has been brought for the non-integer order differential and integral calculus, from
which both the theory and its applications develop significantly. Fractional calculus always remains an interesting, but abstract, mathematical concept up to now \cite{1029},

Fractional calculus has been popularly studied and applied to all kinds of fields, due to its outstanding features, such as long-term memory, non-locality, weak singularity and
capability\cite{1030}. Fractional order systems and controllers have been applied to improve performance and robustness properties in control design\cite{1031}. Non-integer
derivative has been involved in fractional Fourier transform as well as its generalizations, which have gained much popularity in the last a few of decades because of its
numerous applications in signal analysis and optics\cite{1032}, such as filtering, encoding, watermarking, and phase retrieval \cite{1029}. Especially, fractional orthonormal
basis functions are synthesized from generalizing the well-known fixed pole rational basis functions \cite{1029}. A modified Riemann--Liouville definition of fractional
derivative is introduced, which provides a Taylor’s series of fractional order for non differentiable functions, and can act as a framework for a differential geometry of
fractional order \cite{1039}. In \cite{1035}, a scaling laws of human travel was proposed, which is a bifractional diffusion equation. This indicates that the fractional
calculus demonstrates the real changing process of natural phenomena in large possibility. For instance, the fractional Fokker-Planck equation can model the memory
and persistent long-range correlations, as observed widely in neuroscience, such as in the fluctuations of the alpha rhythm\cite{1036}.

Especially, fractional calculus has been taken up in applications in the domain of signal processing and image processing\cite{1033} due to its capability enhancing the textural
detail of signals in a nonlinear manner. For image processing, filters based on fractional derivatives can not only maintain the low-frequency contour features in smooth areas
in a nonlinear fashion, but also can create the possibility enhancing the high-frequency edges and textural details in the areas where the grey level undergoes frequent or
unusual variations, in a nonlinear manner. A fractional-order variational framework was proposed for retinex \cite{1030}, which is a fractional-order partial differential
equation (FPDE) formulation of retinex, and can fulfill the multi-scale nonlocal contrast enhancement with edges and textural details preserved. This is a fundamental important
advantage, making the work proposed in \cite{1030} superior to the filters based on traditional integer-order calculus, especially in images with abundant textures. In
geophysical exploration field, fractional-order calculus theory has also been discussed, seismic records are often contaminated with various kinds of noise, making it very
difficult to distinguish geological features. To this end, a novel adaptive variable time fractional-order anisotropic diffusion equation was introduced \cite{1033}, which can
effectively remove noise, and preserve as well as significantly enhance the coherent seismic events indicating important geological structures.

Based on the aforementioned known results, we can see, most of the works done so far are based on the use of real order fractional calculus, though it is worth to mention that
there are some works modeling phenomena by virtue of complex order fractional derivatives, which also stem from Riemann-Liouville fractional derivatives \cite{1037}\cite{1040}
or from Cauchy and Weyl integrals \cite{1042}. Complex order differ-integrations was visited in \cite{1019}, which showed that the imaginary part is what is difficult to
explain physically at present. About the issue, in \cite{1019} it is thought reasonable that a real order differentiation process can be decomposed or broken into product of two
complex conjugated derivatives. Actually, this explanation does not make much sense. This paper aims at building filters based on complex-order calculus in a new direction, and
the obtained filters demonstrate amazing performance in a variety of applications such as image processing etc. The contributions of this paper can be summarized as follows.
\begin{itemize}
  \item[i:] First, in frequency domain, based on discrete fourier transform, we define the complex-order derivative or integral filter. Especially, a special handling is performed to remove singularities.
  \item[ii:] Second, we build the connection between the binomial distribution of $(1-1)^{\alpha}$ or $(1+1)^{\alpha}$, and the complex $\alpha$-order derivative or integral filter, using multisection theorem. This indicates that, the complex $\alpha$-order derivative or integral $n$-length filter is the superpositions of the $n$-length sections, gotten by orderly partitioning the expansion of $(1-1)^{\alpha}$ or $(1+1)^{\alpha}$.
  \item[iii:] For input signals, the filtered results due to complex-order filters usually have real and imaginary parts. The both as well as their fusions, the phase angle and the modulus parts, can uncover more information concealed in the signal. Several examples have been employed to demonstrate this. It is very astonishing that using complex-order integral filters can indicate the diseased regions in TRUS images, which is against the traditional views using derivative operations to disclose them.
\end{itemize}

This paper is organized as follows: in Section 2, mathematical apparatus for complex-order filters is given. In Section 3, we give the
algorithms how to build the filters, together with numerical results as well as some discussions. Finally, conclusions are outlined in
Section 4. 

\section{Mathematical Apparatus} \label{sec.ma}
Complex-order calculus usually is nonlocal, whose energy is distributed in an infinite-dimensional space. Of course, when the order changes from a usual complex number to an
integer number, the calculus becomes local, and the energy is distributed in a finite-dimensional space naturally. So, in building a filter, there exist a contradiction between
requiring it to be local and requiring it to have usual complex-order (non-integer) calculus. In this section, theoretical foundation is presented for getting over the
contradiction, for building filters with finite dimensions and with usual complex-order calculus.
\subsection{Preliminaries}
Let $\mathbf{X}=[x_{0},x_{1},\ldots,x_{n-1}]^{T}\in \mathcal{R}^{n}$ denote a data sequence, a sub-sequence of $\mathbf{X}$ is denoted as
$\mathbf{X}_{[0:k-1,k+2]}=[x_{0},x_{1},\ldots,x_{k-1},x_{k+2}]^{T}\in \mathcal{R}^{k+1}$, and the data sequence with $\mathbf{X}$'s elements reversed is denoted as
$\widehat{\mathbf{X}}$. Elementwise power is denoted as $\mathbf{X}^\alpha=[x_{0}^{\alpha},x_{1}^{\alpha},\ldots,x_{n-1}^{\alpha}]^{T}$ if there is no special illustration. Let
$\mathcal{F}(\mathbf{X})=[u_{0},u_{1},\ldots,u_{n-1}]^{T}$ denote the Discrete Fourier Transformation (DFT) of $\mathbf{X}$, `$\odot$' elementwise product, and `$\ast$'
convolution operator. Define
\bea \nonumber
\mathbf{D}=[1-\exp(\frac{-2\pi i}{n}j)]_{n\times1},j=0,\ldots,n-1
\eea and
\bea \nonumber
\mathbf{I}=[1+\exp(\frac{-2\pi i}{n}j)]_{n\times1},j=0,\ldots,n-1.
\eea
Let `$\lfloor x \rfloor$' denote an integer not higher than $x$. Let $\mathcal{L}$ denote a vector whose entries are one, and whose dimension are consistent with the context's
requirements. Let $\mathcal{R}_{e}(x)$, $\mathcal{I}_{m}(x)$, $\mathcal{A}_{n}(x)$ and $\mathcal{M}_{o}(x)$ denote the real part, imaginary part, phase angle and modulus of $x$ respectively. Let `$\mathcal{D}^{\alpha}_{n}$' and
`$\mathcal{I}^{\alpha}_{n}$' denote $\alpha$-order derivative and integral operator in $n$-dimension space, respectively; $\mathbf{P}_{n}=[p_{i,j}]_{n\times n}$ $n-$deimensional
circulant matrix whose entries $p_{\text{mod}(k,n)+1,k}=1$, and others zeros. Let $\mathcal{N}(\mathbf{A})$ denote the orthonormal basis for the null space of $\mathbf{A}$, that
is, $\mathbf{A} \mathcal{N}(\mathbf{A})=0$

Before giving the formal discussion, two lemmas are introduced for the sake of making this paper self-explanatory.

 \begin{lemma} \label{lemma1} There exist $\sum_{k=0}^{\infty}\sum_{j=0}^{n-1}\exp(-\frac{2\pi i}{n}(k+h)j)=\sum_{p=0}^{\infty}n\delta(pn-h)$,
 where $\delta(x)$ is the Dirac delta function: $\delta(x)=1$ when $x=0$; $\delta(x)=0$ otherwise.
\end{lemma}
{\it {\bf Proof}} Its proof can be found in textbooks, and omitted here.
\begin{lemma} \label{lemma2} For any complex number $\alpha$, if $\mathcal{R}_{e}(\alpha)\ge0$, there exists
$(1+x)^{\alpha}=\sum_{k=0}^{\infty}(^{\alpha}_{k})x^k
$ for $|x|=1$ and $x\neq -1$.
\end{lemma}
{\it {\bf Proof}} The proof can refer to Theorem 7.46 in \cite{1045}.

\subsection{Mathematical Derivations}
For a data sequence $\mathbf{X}=[x_{0},x_{1},\ldots,x_{n-1}]$, its DFT is
\bea \label{eq.fourier}
u_{k}=\sum_{j=0}^{n-1}\exp(-\frac{2\pi i}{n}kj)x_{j}.
\eea
For the data sequence $\mathcal{D}^{1}_{n}\mathbf{X}=[x_{0},x_{1}-x_{0},x_{2}-x_{1},\ldots,x_{n-1}-x_{n-2}]$, its DFT is
\bea \nonumber
&& v_{k}=\exp(-\frac{2\pi i}{n}k\times 0)x_{0}+
\\ \nonumber &&\hspace{0.8cm}
\sum_{j=1}^{n-1}\exp(-\frac{2\pi i}{n}kj)(x_{j}-x_{j-1})
\\ \nonumber &&\hspace{0.4cm}=\sum_{j=0}^{n-2}\exp(-\frac{2\pi i}{n}kj)x_{j}[1-\exp(-\frac{2\pi i}{n}k)]
\\ \nonumber &&\hspace{0.6cm}+x_{n-1}\exp(-\frac{2\pi i}{n}k(n-1))
\eea Substituting \eqref{eq.fourier} into above formula gives that
\bea \nonumber
&&v_{k}=[u_{k}-x_{n-1}\exp(-\frac{2\pi i}{n}k(n-1))]
\\ \nonumber &&\hspace{0.6cm}[1-\exp(-\frac{2\pi i}{n}k)]+x_{n-1}\exp(-\frac{2\pi i}{n}k(n-1))
\\ \nonumber &&\hspace{0.4cm}=u_{k}[1-\exp(-\frac{2\pi i}{n}k)]+
\\ \nonumber &&\hspace{0.6cm}x_{n-1}\exp(-\frac{2\pi i}{n}k(n-1))\exp(-\frac{2\pi i}{n}k)
\\ \label{eq.fourier.ex1} &&\hspace{0.4cm}=u_{k}[1-\exp(-\frac{2\pi i}{n}k)]+x_{n-1}
\eea
From \eqref{eq.fourier} and \eqref{eq.fourier.ex1}, we can observe that
\bea \label{eq.adriv}
\mathcal{D}^{1}_{n}\mathbf{X}=\mathcal{F}^{-1}(\mathcal{F}(\mathbf{X})\odot\mathbf{D}^{1})=\mathcal{F}^{-1}(\mathbf{D}^{1})*\mathbf{X}
\eea
In analogy with \eqref{eq.adriv}, we can also have
\bea \label{eq.ainteg}
\mathcal{I}^{1}_{n}\mathbf{X}=\mathcal{F}^{-1}(\mathbf{I}^{1})*\mathbf{X}
\eea

From \eqref{eq.adriv} and \eqref{eq.ainteg}, we can take it for granted that the operator $\mathcal{D}^{1}_{n}$ and $\mathcal{I}^{1}_{n}$ correspond to
$\mathcal{F}^{-1}(\mathbf{D}^{1}))$ and $\mathcal{F}^{-1}(\mathbf{I}^{1}))$ respectively. Can we replace `1' in \eqref{eq.adriv} and \eqref{eq.ainteg} with any positive
integer $m$? For this generalization, it is natural indeed, due to the following relations.
\bea \label{eq.mdriv}
\mathcal{D}^{m}_{n}\mathbf{X}=\underbrace{\mathcal{F}^{-1}(\mathbf{D}^{1})*(...(\mathcal{F}^{-1}(\mathbf{D}^{1})}_{m}*\mathbf{X})=\mathcal{F}^{-1}(\mathbf{D}^{m})*\mathbf{X},
\eea
\bea \label{eq.mInteg}
\mathcal{I}^{m}_{n}\mathbf{X}=\underbrace{\mathcal{F}^{-1}(\mathbf{I}^{1})*(...(\mathcal{F}^{-1}(\mathbf{I}^{1})}_{m}*\mathbf{X})=\mathcal{F}^{-1}(\mathbf{I}^{m})*\mathbf{X}.
\eea
Can we generalize further, replacing $m$ with a non-integer, even usual complex, number $\alpha$? In the following, we will mainly concentrate on discussing the properties
$\mathcal{F}^{-1}(\mathbf{D}^{\alpha})$ as well as $\mathcal{F}^{-1}(\mathbf{I}^{\alpha})$, with $\alpha$ being an usual complex number, and try to answer the physical meaning
of
raised in \cite{1019}. As indicated \eqref{eq.mdriv} and \eqref{eq.mInteg}, $\mathcal{F}^{-1}(\mathbf{D}^{\alpha})$ and $\mathcal{F}^{-1}(\mathbf{I}^{\alpha})$ can be seen as
$\alpha$-order derivative and integral operators, $\mathcal{D}^{\alpha}_{n}$ and $\mathcal{I}^{\alpha}_{n}$, respectively. Specially, to remove the singularity of $\mathcal{D}^{\alpha}_{n}$
due to the first element, `0', in $\mathbf{D}$, we let
\bea \label{eq.drv.real}
\mathcal{D}^{\alpha}_{n}=\mathcal{F}^{-1}([0;\mathbf{D}_{[1:n-1]}^{\alpha}]).
\eea Similarly, when $n$ is even, let
\bea \label{eq.int.real}
\mathcal{I}^{\alpha}_{n}=\mathcal{F}^{-1}([\mathbf{I}_{[0:n/2-1]}^{\alpha};0;\mathbf{I}_{[n/2+1:n]}^{\alpha}]).
\eea
By these tricks, for the derivative and integral operators, $\mathcal{D}^{\alpha}_{n}$ and $\mathcal{I}^{\alpha}_{n}$, the feasible domain of $\alpha$ is extended, and fills up the
complex domain without singularity.

To understand $\mathcal{D}^{\alpha}_{n}$ and $\mathcal{I}^{\alpha}_{n}$, some properties about them are given in the form of four theorems as well as the associated remarks. Two theorems
are for real number $\alpha$, and the other two are for usual complex number $\alpha$.

\begin{theorem} \label{theorem2} The elements of $\mathcal{D}^{\alpha}_{n}$ and $\mathcal{I}^{\alpha}_{n}$ are all real numbers, when $\alpha$ is a real number.
\end{theorem}

 {\it {\bf Proof}}: Let
 \bea \label{eq.Dna}
\mathcal{D}^{\alpha}_{n}\equiv[x_{0},x_{1},\ldots,x_{n-1}]^{T},
\\ \label{eq.Ina}
\mathcal{I}^{\alpha}_{n}\equiv[y_{0},y_{1},\ldots,y_{n-1}]^{T}.
\eea
Using Inverse Discrete Fourier Transform and considering \eqref{eq.drv.real}, we know
\bea \nonumber
 && x_{k}=0+\sum_{j=1}^{n-1}\exp(\frac{2\pi i}{n} kj)(1-\exp(-\frac{2\pi i}{n}j) )^{\alpha}
 \\ \nonumber
 &&\hspace{0.4cm}=\sum_{j=1,2j< n}^{n-1}\left(\exp(\frac{2\pi i}{n} kj)(1-\exp(-\frac{2\pi i}{n}j) )^{\alpha}\right.
 \\ \nonumber
 &&\hspace{0.6cm}\left.+\exp(\frac{2\pi i}{n} k(n-j))(1-\exp(-\frac{2\pi i}{n}(n-j)) )^{\alpha}\right)
  \\ \nonumber
 &&\hspace{0.6cm}+(1-\text{mod}(n,2))\exp(\frac{2\pi i}{n} k\frac{n}{2})(1-\exp(-\frac{2\pi i}{n}\frac{n}{2}) )^{\alpha}
 \\ \nonumber
 &&\hspace{0.4cm}=\sum_{j=1}^{\lfloor(n-1)/2\rfloor}\left(\exp(\frac{2\pi i}{n} kj)(1-\exp(-\frac{2\pi i}{n}j) )^{\alpha}\right.
 \\ \label{eq.1}
 &&\hspace{0.6cm}\left.+\exp(-\frac{2\pi i}{n} kj)(1-\exp(\frac{2\pi i}{n}j) )^{\alpha}\right)
  \\ \nonumber
 &&\hspace{0.6cm}+(1-\text{mod}(n,2))(-1)^{k}2^{\alpha}.
  \eea Similarly we can derive
  \bea \nonumber
 && y_{k}=2^{\alpha}+\sum_{j=1}^{\lfloor(n-1)/2\rfloor}\left(\exp(\frac{2\pi i}{n} kj)(1+\exp(-\frac{2\pi i}{n}j) )^{\alpha}\right.
 \\ \label{eq.2}
 &&\hspace{0.6cm}\left.+\exp(-\frac{2\pi i}{n} kj)(1+\exp(\frac{2\pi i}{n}j) )^{\alpha}\right)
  \eea
when $\alpha$ is a real number, in \eqref{eq.1} the item $\exp(\frac{2\pi i}{n} kj)(1-\exp(-\frac{2\pi i}{n}j) )^{\alpha}$ is conjugate to $\exp(-\frac{2\pi i}{n}
kj)(1-\exp(\frac{2\pi i}{n}j) )^{\alpha}$, and in \eqref{eq.2} the item $\exp(\frac{2\pi i}{n} kj)(1+\exp(-\frac{2\pi i}{n}j) )^{\alpha}$ is conjugate to $\exp(-\frac{2\pi i}{n}
kj)(1+\exp(\frac{2\pi i}{n}j) )^{\alpha}$. Because the sum of two items is a real number when the two items are conjugate to each other, we can conclude that $x_{k}$ and $y_{k}$
are all real numbers. This completes the proof. $\blacksquare$

 \begin{theorem} \label{theorem2.1} If $\alpha$ is an positive integer $m$, then $\mathcal{D}^{m}_{n}$ and $\mathcal{I}^{m}_{n}$ are integer $m$-order difference and integral
 masks, which are actually binomial coefficients of $(1-1)^{m}$ and $(1+1)^{m}$, respectively. When $m<n$, the coefficients are orderly placed on dimensions of the
 $n$-dimensional space, otherwise, the coefficients are superposed within the $n$-dimensional space.
\end{theorem}

 {\it {\bf Proof}}: Based on \eqref{eq.Dna} and \eqref{eq.Ina}, from \eqref{eq.1} we know there exists
 \bea \nonumber
 &&x_{k}=\sum_{j=0}^{n-1}\exp(\frac{2kj}{n} \pi i)(1-\exp(-\frac{2\pi i}{n}j) )^{m}
 \\ \nonumber &&\hspace{0.4cm} =\sum_{j=0}^{n-1}\exp(\frac{2kj}{n}\pi i )\sum_{h=0}^{m}(_{h}^{m})(-1)^{h}\exp(-\frac{2 hj }{n}\pi i)
 \\ \label{eq.th2.1} &&\hspace{0.4cm} =\sum_{h=0}^{m}(_{h}^{m})(-1)^{h}\sum_{j=0}^{n-1}\exp(-\frac{2(h-k)j}{n}\pi i ).
  \eea In terms of Lemma \ref{lemma1}, there exists
  \bea \label{eq.11}
  \sum_{j=0}^{n-1}\exp(\frac{-2(h-k)j}{n}\pi i )=n \delta(h-k-pn), p\in \mathbf{Z}
  \eea which indicates that if the difference between $h$ and $k$ is $n$-periodic, $\sum_{j=0}^{n-1}\exp(-\frac{2(h-k)j}{n}\pi i )$ is $n$, otherwise is zero.

  Substituting \eqref{eq.11} into \eqref{eq.th2.1} gives that
  \bea \label{eq.xk.m}
 &&x_{k}=\sum_{h=0}^{m}(_{h}^{m})(-1)^{h}n \delta(h-k-pn)
 \\ \nonumber &&\hspace{0.4cm} =\sum_{p=0,1,\ldots}^{0 \le k+pn\le m}n(_{k+pn}^{m})(-1)^{k+pn}.
  \eea Similarly, we can get
  \bea \label{eq.yk.m}
  y_{k}=\sum_{p=0,1,\ldots}^{0 \le k+pn\le m}n(_{k+pn}^{m}).
 \eea
From \eqref{eq.xk.m} and \eqref{eq.yk.m}, we can see, when $m<n$, $\mathcal{D}^{m}_{n}$ and $\mathcal{I}^{m}_{n}$ are given below
\bea \label{eq.drv.mln.expand}
\mathcal{D}^{m}_{n}=n\left[\underbrace{(_{0}^{m})(-1)^{0},(_{1}^{m})(-1)^{1},\ldots,(_{m}^{m})(-1)^{m}}_{m+1},\underbrace{0,\ldots,0}_{n-(m+1)}\right]^{T}
\eea
\bea\label{eq.int.mln.expand}
\mathcal{I}^{m}_{n}=n\left[\underbrace{(_{0}^{m}),(_{1}^{m}),\ldots,(_{m}^{m})}_{m+1},\underbrace{0,\ldots,0}_{n-(m+1)}\right]^{T} \hspace{2.3cm}
\eea
 From \eqref{eq.drv.mln.expand} and \eqref{eq.int.mln.expand}, we can see the ahead $m+1$ elements of $\mathcal{D}^{m}_{n}$ and $\mathcal{I}^{m}_{n}$ are coincidentally being
 the coefficients of $(1-1)^{m}$ and $(1+1)^{m}$, respectively. Moreover, all the coefficients are not superposed. But when $m \geq n$, what will happen? To intuitively
 demonstrate this issue, we take the expansion of $\mathcal{I}^{m}_{n}$ in terms of \eqref{eq.yk.m} as an example.
 \bea \nonumber
\mathcal{I}^{m}_{n}=n\left[(_{0}^{m})+(_{n}^{m})+\ldots,(_{1}^{m})+(_{n+1}^{m})+\ldots,\ldots,\right.
\\ \label{eq.mgn.int}
\left.(_{n-1}^{m})+(_{2n-1}^{m})+\ldots\right]^{T},
 \eea
which clearly shows that when $m \geq n$ holds, the expansion coefficients of $(1+1)^{m}$ are $n$-periodically folded in the $n$-dimensional space. The expansion of
$\mathcal{D}^{m}_{n}$ for $m \geq n$ can be similarly gotten. So, $\mathcal{D}^{m}_{n}$ and $\mathcal{I}^{m}_{n}$ have the binomial coefficients of $(1-1)^{m}$ and $(1+1)^{m}$
in essence. The sets of the coefficients of $(1-1)^{m}$ and $(1+1)^{m}$ are the classical difference and integral filters respectively, thus $\mathcal{D}^{m}_{n}$ and
$\mathcal{I}^{m}_{n}$ can precisely serve as integer $m$-order difference and integral masks. This completes the proof.  $\blacksquare$

Theorem \ref{theorem2.1} actually constructs the connection between DFT and binomial expansion, and the coefficients of binomial expansion are related to difference and integral
operations. So, theorem \ref{theorem2.1} indicates 1) the $m$th element-wise power of $\mathbf{D}$ and $\mathbf{I}$ in frequency domain corresponds to $m$-order difference and
integral filters in spatial space; and 2) In making $\mathcal{D}^{m}_{n}$ and $\mathcal{I}^{m}_{n}$ come into being, the spatial energy corresponding to the $m$th element-wise
power of $\mathbf{D}$ and $\mathbf{I}$, may be superposed on the $n$ coordinates of the $n$-dimensional spatial space, depending on the value relationship of $m$ and $n$. In
all, about Theorem \ref{theorem2.1} we have the following remark.

\begin{remark} \label{rmk1} For a positive integer $m$, the $n$-length filters $\mathcal{D}^{m}_{n}$ and $\mathcal{I}^{m}_{n}$ actually are the superpositions of the $m+1$
binomial coefficients of $(1-1)^{m}$ and $(1+1)^{m}$.
\end{remark}

When $\alpha$ is a usual complex number, that is to say, when $\alpha$ is not an inter number, even not a real number, how are $\mathcal{D}^{m}_{n}$ and $\mathcal{I}^{m}_{n}$ to
be? About this question, we have the following acquaintances with two theorems and the associated remarks given.

 \begin{theorem} \label{theorem3} For any complex number $\alpha\in \mathcal{C}$, there exist $\mathbf{L}^{T}\mathcal{D}^{\alpha}_{n}=0$ and $\mathbf{L}^{T}
 \mathcal{I}^{\alpha}_{n}=n2^\alpha$.
\end{theorem}
{\it {\bf Proof}}: In terms of Lemma \ref{lemma1}, we have
 \bea \nonumber
 &&\mathbf{L}^{T}\mathcal{D}^{\alpha}_{n}=\sum_{k=0}^{n-1}\sum_{j=1}^{n-1}\exp(\frac{2kj}{n} \pi i)(1-\exp(-\frac{2\pi i}{n}j) )^{\alpha}
 \\ \nonumber &&\hspace{1.1cm} =\sum_{j=1}^{n-1}\sum_{k=0}^{n-1}\exp(\frac{2kj}{n} \pi i)(1-\exp(-\frac{2\pi i}{n}j) )^{\alpha}
 \\ \nonumber &&\hspace{1.1cm} =\sum_{j=1}^{n-1}n\delta(j)(1-\exp(-\frac{2\pi i}{n}j) )^{\alpha}
 \\ \nonumber &&\hspace{1.1cm} =n(1-\exp(-\frac{2\pi i}{n}0) )^{\alpha}
 \\ \label{tho3.eq1} &&\hspace{1.1cm} =0.
  \eea Similarly, we get
 \bea \nonumber
 &&\mathbf{L}^{T} \mathcal{I}^{\alpha}_{n}=n(1+\exp(-\frac{2\pi i}{n}0) )^{\alpha}
 \\ \label{tho3.eq2} &&\hspace{1.0cm} =n2^{\alpha}.
  \eea
  This completes the proof. $\blacksquare$

For any usual complex number $\alpha$, Theorem \ref{theorem3} demonstrates that the real and imaginary parts of $\mathcal{D}^{\alpha}_{n}$ are $0$. That is to say,
$\mathcal{R}_{e}(\mathcal{D}^{\alpha}_{n})$ and $\mathcal{I}_{m}(\mathcal{D}^{\alpha}_{n})$ can serve as derivative filters. On the other side,
$\mathcal{I}^{\alpha}_{n}=n2^{\alpha}$ makes
$\mathcal{I}^{\alpha}_{n}$ have the potential being a smooth filter, like that the coefficients of $(1+1)^{m}$ forms an $m$-order smooth filter with length $m+1$.

\begin{theorem} \label{theorem4} For any $\alpha \in \mathbf{C}$, there exist
\bea \nonumber
\mathcal{D}^{\alpha}_{n}=n\left[\sum_{p=0}^{\infty}(_{0+pn}^{\alpha})(-1)^{0+pn},\sum_{p=0}^{\infty}(_{1+pn}^{\alpha})(-1)^{1+pn}\right.
\\ \label{eq.drv.nnegalpha}
\left.\ldots,\sum_{p=0}^{\infty}(_{n-1+pn}^{\alpha})(-1)^{n-1+pn}\right]^{T},\mathcal{R}_{e}(\alpha)> 0
\eea
and
\bea \label{eq.drv.negalpha}
\mathcal{D}^{\alpha}_{n}=\frac{-\widehat{\mathcal{N}(\mathbf{M}^{-\alpha}_{n})}}{-n\mathcal{N}^{T}(\mathbf{M}^{-\alpha}_{n})\mathcal{D}^{-\alpha}_{n}},\mathcal{R}_{e}(\alpha)\le 0;
\eea as well as
\bea \label{eq.int.evenn.negalpha}
\mathcal{I}^{\alpha}_{n}=\frac{1}{n}(\mathbf{S}^{-\alpha}_{n})^{T}(\mathbf{S}^{-\alpha}_{n}(\mathbf{S}^{-\alpha}_{n})^{T})^{-1}[1,-1,1,-1,\ldots,1]_{1\times n}^{T}
\eea for even $n$ and $\mathcal{R}_{e}(\alpha) \le 0$, and
\bea
\label{eq.int.others}
\mathcal{I}^{\alpha}_{n}=n\left[\sum_{p=0}^{\infty}(_{0+pn}^{\alpha}),\sum_{p=0}^{\infty}(_{1+pn}^{\alpha}),\ldots,\sum_{p=0}^{\infty}(_{n-1+pn}^{\alpha})\right]^{T}
\eea for others,
where
\bea \nonumber
\mathbf{M}^{-\alpha}_{n} \equiv \left[
                                  \begin{array}{c}
                                    \mathcal{D}^{-\alpha}_{n}(\mathbf{P}_{n}^{3}-\mathbf{P}_{n}^{2}) \\
                                    \vdots \\
                                    \mathcal{D}^{-\alpha}_{n}(\mathbf{P}_{n}^{n}-\mathbf{P}_{n}^{n-1}) \\
                                    \mathbf{L} \\
                                  \end{array}
                                \right],
                                \mathbf{S}_{n}^{-\alpha}\equiv\left[
      \begin{array}{c}
        \mathcal{I}^{-\alpha}_{n}\mathbf{P}_{n}^{2} \\
        \ldots \\
        \mathcal{I}^{-\alpha}_{n}\mathbf{P}_{n}^{n} \\
      \end{array}
    \right].
\eea
\end{theorem}
 {\it {\bf Proof}}: Based on \eqref{eq.Dna} and \eqref{eq.Ina}, in terms of Lemma \ref{lemma2}, when $\mathcal{R}_{e}(\alpha)\ge 0$, like the derivations from \eqref{eq.th2.1}
 to \eqref{eq.xk.m}, we have
 \bea \nonumber
&&x_{k}=\sum_{j=0}^{n-1}\exp(\frac{2kj}{n} \pi i)(1-\exp(-\frac{2\pi i}{n}j) )^{\alpha}
\\ \label{eq14}
&&\hspace{0.4cm} =n\sum_{p=0}^{\infty}(_{k+pn}^{\alpha})(-1)^{k+pn}
 \eea
 and
\bea \label{eq15}
y_{k}=n\sum_{p=0}^{\infty}(_{k+pn}^{\alpha}).
\eea
When $\mathcal{R}_{e}(\alpha)< 0$, $x_{k}$ and $y_{k}$ cannot be obtained from \eqref{eq14} and \eqref{eq15} as Lemma \ref{lemma2} does not hold. To get $x_{k}$ and $y_{k}$ for
$\mathcal{R}_{e}(\alpha)< 0$, we use the convolution rule and Theorem \ref{theorem3}. In terms of \ref{eq.drv.real}, we know there exists
\bea \nonumber
&&\mathcal{D}^{-\alpha}_{n}*\mathcal{D}^{\alpha}_{n}=\mathcal{F}^{-1}([0,\mathbf{D}_{[1:n-1]}^{-\alpha}]\odot[0,\mathbf{D}_{[1:n-1]}^{\alpha}])
\\ \nonumber
&&\hspace{1.6cm}=\mathcal{F}^{-1}([0,\underbrace{1,\ldots,1}_{n-1}])
\\ \label{eq16}
&&\hspace{1.6cm}=\frac{1}{n}[n-1,\underbrace{-1,\ldots,-1}_{n-1}].
\eea
Due to Theorem \ref{theorem3}, in \eqref{eq16} there are only $n-1$ constraints
\bea \label{eq17}
(\mathcal{D}^{-\alpha}_{n})^{T}\mathbf{P}^n_{n}\widehat{\mathcal{D}}^{\alpha}_{n}=(\mathcal{D}^{-\alpha}_{n})^{T}\widehat{\mathcal{D}}^{\alpha}_{n}=-\frac{1}{n}.
\eea and
\bea \label{eq18}
(\mathcal{D}^{-\alpha}_{n})^{T}\mathbf{P}^2_{n}\widehat{\mathcal{D}}^{\alpha}_{n}=(\mathcal{D}^{-\alpha}_{n})^{T}\mathbf{P}^3_{n}\widehat{\mathcal{D}}^{\alpha}_{n}=\ldots=(\mathcal{D}^{-\alpha}_{n})^{T}\mathbf{P}^n_{n}\widehat{\mathcal{D}}^{\alpha}_{n}
\eea
The effect of constraint \eqref{eq17} is only to perform normalization, and from \eqref{eq18} we can get
\bea \label{eq19}
(\mathcal{D}^{-\alpha}_{n})^{T}(\mathbf{P}^k_{n}-\mathbf{P}^{k-1}_{n})\widehat{\mathcal{D}}^{\alpha}_{n}=0 \text{ for } k=3,\ldots,n.
\eea
Due to \eqref{tho3.eq1}, we know
\bea \label{eq20}
\mathbf{L}^{T} \widehat{\mathcal{D}}^{\alpha}_{n}=0.
\eea
Combining \eqref{eq19} and \eqref{eq20}, we know
\bea
\mathbf{M}^{-\alpha}_{n}\widehat{\mathcal{D}}^{\alpha}_{n}=0,
\eea
which indicates that $\widehat{\mathcal{D}}^{\alpha}_{n}$ belongs to the null space of $\mathbf{M}^{-\alpha}_{n}$, $\mathcal{N}(\mathbf{M}^{-\alpha}_{n})$, which is a
one-dimensional space. In terms of the normalization procedure \eqref{eq17}, the final $\mathcal{D}^{\alpha}_{n}$ can be gotten, as shown in \eqref{eq.drv.negalpha}.

For even $n$ and $\mathcal{R}_{e}(\alpha)< 0$, based on \eqref{eq.int.real}, like \eqref{eq16} we have
\bea \nonumber
&&\mathcal{I}^{-\alpha}_{n}*\mathcal{I}^{\alpha}_{n}=\mathcal{F}^{-1}([\underbrace{1,\ldots,1}_{n/2},0,[\underbrace{1,\ldots,1}_{n/2-1}])
\\ \nonumber
&&\hspace{1.6cm}=\frac{1}{n}[n-1,\underbrace{1,-1,1,-1,\ldots,1}_{n-1}],
\eea which indicates
\bea \label{eq21}
\left[
      \begin{array}{c}
        \mathcal{I}^{-\alpha}_{n}\mathbf{P}_{n}^{1} \\
        \mathcal{I}^{-\alpha}_{n}\mathbf{P}_{n}^{2} \\
        \ldots \\
        \mathcal{I}^{-\alpha}_{n}\mathbf{P}_{n}^{n} \\
      \end{array}
    \right]\mathcal{I}^{\alpha}_{n}=\frac{1}{n}[n-1,\underbrace{1,-1,1,-1,\ldots,1}_{n-1}]^{T}.
\eea
The equation set \eqref{eq21} only has $n-1$ independent constraints, and also has the constraint specified in \eqref{tho3.eq2}, so integrating \eqref{tho3.eq2} and \eqref{eq21} gives that
\bea
\mathbf{S}_{n}^{-\alpha}\mathcal{I}^{\alpha}_{n}=\frac{1}{n}[1,-1,1,-1,\ldots,1]_{1\times (n-1)}^{T},
\eea from which, \eqref{eq.int.evenn.negalpha} is derived. This completes the proof. $\blacksquare$

The equations \eqref{eq.drv.nnegalpha} and \eqref{eq.int.others} actually can be seen as the extensions of \eqref{eq.xk.m} and \eqref{eq.yk.m} accordingly, and are the results by n-periodically folding the coefficients of the expansions of $(1-1)^{\alpha}$ and $(1+1)^{\alpha}$, into an $n$-dimension space. Of course, \eqref{eq.xk.m} and \eqref{eq.yk.m} are the special cases of \eqref{eq.drv.nnegalpha} and \eqref{eq.int.others} when $\alpha$ takes positive integers. In summation, \eqref{eq.drv.nnegalpha} and \eqref{eq.int.others} are consistent with the usual binomial distribution difference and integration filtering operations. But, how are \eqref{eq.drv.negalpha} and \eqref{eq.int.evenn.negalpha}? The both are for handling the singularity $0^{\alpha}$ when $\alpha$ has a non-positive real part, and in the cases $0^{\alpha}$ has been replaced with $0$, as shown in \eqref{eq.drv.real} and \eqref{eq.int.real}. This manual intervention has made $\mathcal{D}^{\alpha}_{n}$ and $\mathcal{I}^{\alpha}_{n}$ in the cases not consistent with the conclusion, that $\mathcal{D}^{\alpha}_{n}$ and $\mathcal{I}^{\alpha}_{n}$ are gotten by n-periodically folding the coefficients of the expansions of $(1-1)^{\alpha}$ and $(1+1)^{\alpha}$, as shown in \eqref{eq.drv.nnegalpha} and \eqref{eq.int.others}. But what are the meanings of the entries of $\mathcal{D}^{\alpha}_{n}$ and $\mathcal{I}^{\alpha}_{n}$, shown in \eqref{eq.drv.negalpha} and \eqref{eq.int.evenn.negalpha}? The equation \eqref{eq.drv.negalpha} shows that $\mathcal{D}^{\alpha}_{n}$ are in the null space of $\mathbf{M}_{n}^{-\alpha}$, which originates from $\mathcal{D}^{-\alpha}_{n}$; and \eqref{eq.int.evenn.negalpha} shows that $\mathcal{I}^{\alpha}_{n}$ belongs to space of $\mathbf{S}_{n}^{-\alpha}$. As for the precise meaning of each entry in $\mathcal{D}^{\alpha}_{n}$ and $\mathcal{I}^{\alpha}_{n}$ indicated in \eqref{eq.drv.negalpha} and
\eqref{eq.int.evenn.negalpha}, it seems that the two equations are not enough to imply, because the both do not clearly show the solution for each entry of  $\mathcal{D}^{\alpha}_{n}$ and $\mathcal{I}^{\alpha}_{n}$, like \eqref{eq.drv.nnegalpha} and \eqref{eq.int.others}. Referring to Theorem \ref{theorem3}, combining \eqref{eq.drv.nnegalpha}$\sim$\eqref{eq.int.others} tells that $\mathcal{D}^{\alpha}_{n}$ and $\mathcal{I}^{\alpha}_{n}$ exist for any complex $\alpha$, and can be used for filtering.

\section{The Algorithm and Experimental Test} \label{sec.ma}
\begin{figure}[htbp]
  \centering
   \subfigure[The entries of $\mathcal{D}^{\alpha}_{8}$]{
                \centering
        \includegraphics[width=4cm]{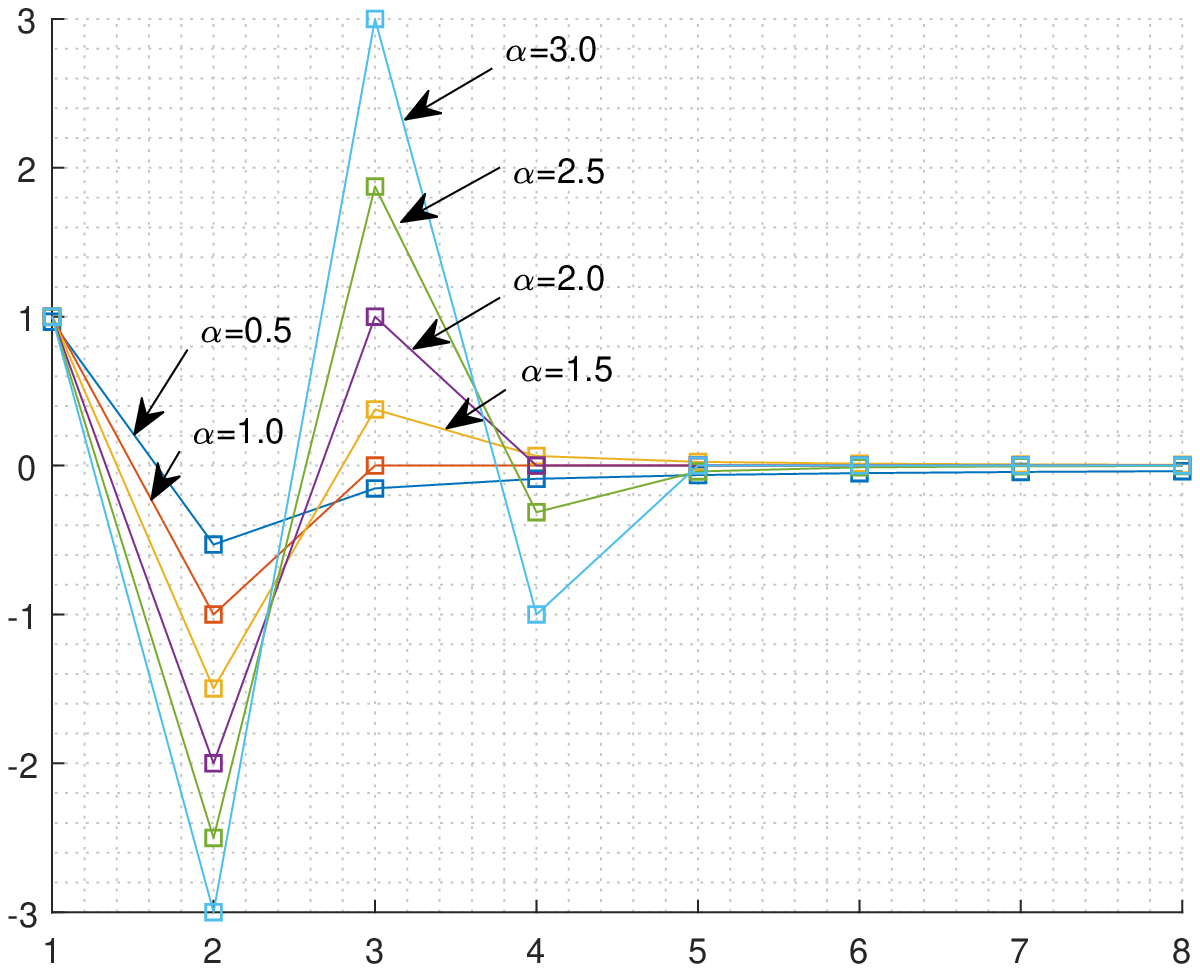}
}
   \subfigure[The entries of $\mathcal{I}^{\alpha}_{8}$]{
                            \centering
        \includegraphics[width=4cm]{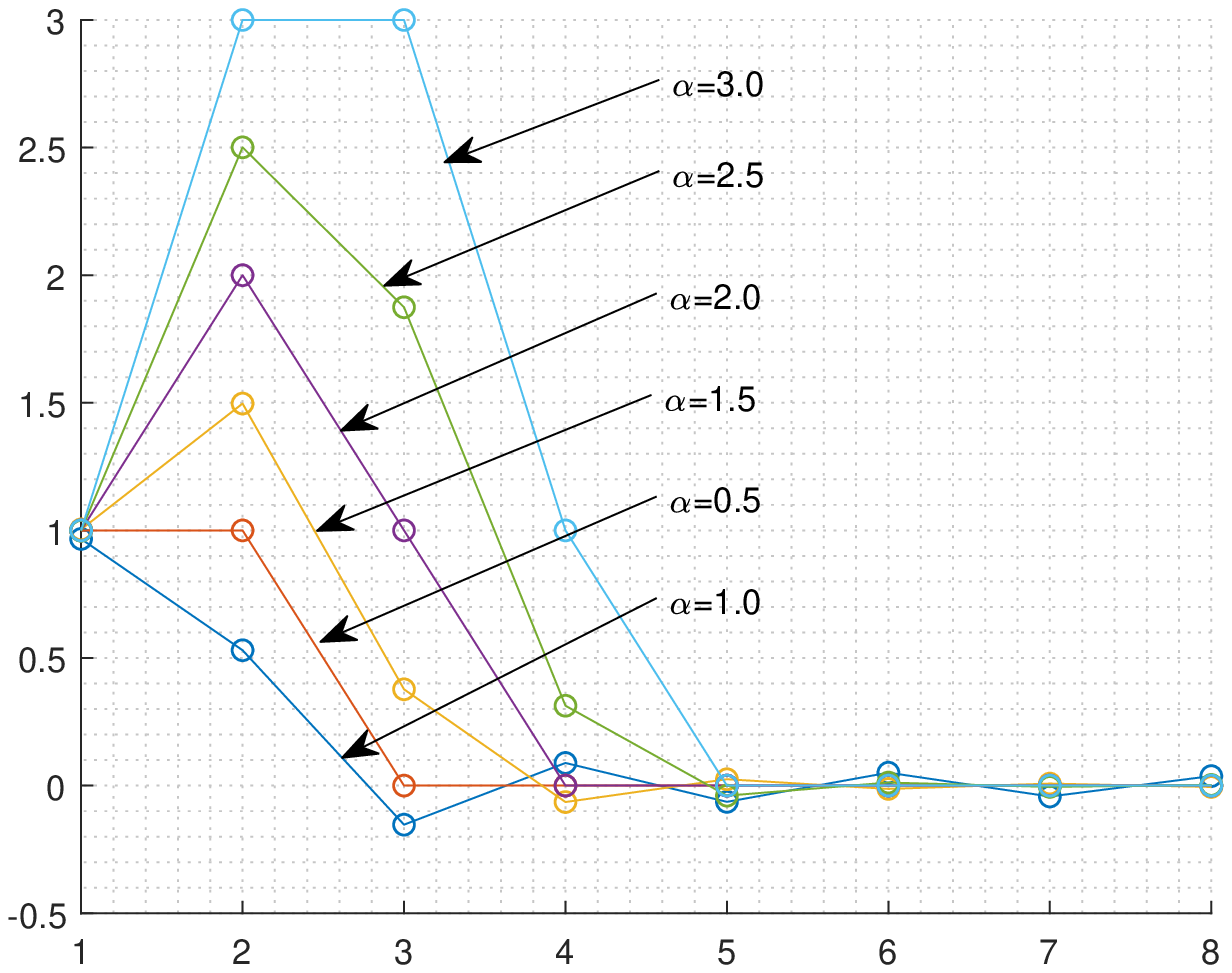}
}
  \caption{The distribution center (if exists) of the non-zero entries of $\mathcal{D}^{\alpha}_{n}$ and $\mathcal{I}^{\alpha}_{n}$ usually is not at the spatial center of $\mathcal{D}^{\alpha}_{n}$ and $\mathcal{I}^{\alpha}_{n}$.}\label{Fig.alpha}
\end{figure}

As is well known, for a filter, its central position is the filtering reference point, and its entries closer to the reference point ought to make more contribution to the filtered consequence. That is to say, entries closer to filtering reference points ought to have larger modules. The filters, $\mathcal{D}^{\alpha}_{n}$ or $\mathcal{I}^{\alpha}_{n}$, produced by \eqref{eq.drv.nnegalpha}$\sim$\eqref{eq.int.others}, even \eqref{eq.xk.m} and \eqref{eq.yk.m}, usually cannot fit in with above rule, as shown in Figure \ref{Fig.alpha}. So $\mathcal{D}^{\alpha}_{n}$ or $\mathcal{I}^{\alpha}_{n}$ cannot be used straightforwardly, and we first need to rearrange the entries of $\mathcal{D}^{\alpha}_{n}$ or $\mathcal{I}^{\alpha}_{n}$. After designing the filters, several instances have been employed to demonstrate the performance.

\subsection{Designing Filters}
When $\alpha$ takes an positive integer $m$, if $m<n$, the nonzero entries of $\mathcal{D}^{\alpha}_{n}$ and $\mathcal{I}^{\alpha}_{n}$ naturally have a distribution center, around at the $\lfloor\frac{m+1}{2}\rfloor$th entry. In this case, what we need to do is to move the distribution center to the central position of $\mathcal{D}^{\alpha}_{n}$ and $\mathcal{I}^{\alpha}_{n}$, around at $\lfloor\frac{n}{2}\rfloor$th entry. If $m\ge n$, according to \eqref{eq.xk.m} and \eqref{eq.yk.m}, binomial expansion can not be fully placed in $n$-dimension space, and superposition happens. In this case, usually there is no modulus distribution center in $\mathcal{D}^{\alpha}_{n}$ and $\mathcal{I}^{\alpha}_{n}$. Of course, this phenomenon will happen when $\alpha$ takes usual complex numbers. In all, the entries in $\mathcal{D}^{\alpha}_{n}$ and $\mathcal{I}^{\alpha}_{n}$ need to be rearranged in most cases.

In terms of the rule that entries closer to the filtering reference point should make more contributions to filtered consequence, the following steps are proposed for designing $q$-dimensional $n_{1}\times n_{2}\ldots\times n_{q}$ hypercube filters $\mathcal{D}^{\alpha}_{n_{1}\times n_{2}\ldots\times n_{q}}$ and $\mathcal{I}^{\alpha}_{n_{1}\times n_{2}\ldots\times n_{q}}$.
\begin{enumerate}
  \item [step 1:] In $q$-dimensional space, let $\left[\lfloor\frac{n_{1}}{2}\rfloor;\ldots;\lfloor\frac{n_{q}}{2}\rfloor\right]_{q\times 1}$ be the filtering reference point;
  \item [step 2:] In one dimensional space, using \eqref{eq.drv.nnegalpha}$\sim$\eqref{eq.int.others} as well as \eqref{eq.xk.m} and \eqref{eq.yk.m} to produce $\mathcal{D}^{\alpha}_{n_{1}\ldots n_{q}}$ and $\mathcal{I}^{\alpha}_{n_{1}\ldots n_{q}}$, they all have $n_{1}\ldots n_{q}$ entries;
  \item [step 3:] According to the entry moduli, place the entries of $\mathcal{D}^{\alpha}_{n_{1}\ldots n_{q}}$ and $\mathcal{I}^{\alpha}_{n_{1}\ldots n_{q}}$ into the hypercube data block of $\mathcal{D}^{\alpha}_{n_{1}\times n_{2}\ldots\times n_{q}}$ and $\mathcal{I}^{\alpha}_{n_{1}\times n_{2}\ldots\times n_{q}}$, the entries with larger moduli ought to be placed closer to the center of the block, $\left[\lfloor\frac{n_{1}}{2}\rfloor;\ldots;\lfloor\frac{n_{q}}{2}\rfloor\right]_{q\times 1}$.
\end{enumerate}

Through step 1-3, $\mathcal{D}^{\alpha}_{n_{1}\times n_{2}\ldots\times n_{q}}$ and $\mathcal{I}^{\alpha}_{n_{1}\times n_{2}\ldots\times n_{q}}$ can be used for filtering. For a data sequence $\mathbf{X}$, whether its entries are real or complex numbers, usually the filtered consequence can reveal more details. The real/imaginary/angle/modulus information of the consequence $\mathcal{D}^{\alpha}_{n_{1}\times n_{2}\ldots\times n_{q}}\mathbf{X}$ has potential usefulness. Compared with $\mathcal{D}^{\alpha}_{n_{1}\times n_{2}\ldots\times n_{q}}$ and $\mathcal{I}^{\alpha}_{n_{1}\times n_{2}\ldots\times n_{q}}$ restricting $\alpha$ to fractions \cite{1019}, the imaginary and angle information is newly generated due to the nonzero imaginary part of $\alpha$. In practice, we can fuse all the information in terms of goals.
\subsection{Experimental Tests and Discussions}
To show the performance of $\mathcal{D}^{\alpha}_{n_{1}\times n_{2}\ldots\times n_{q}}$ and $\mathcal{I}^{\alpha}_{n_{1}\times n_{2}\ldots\times n_{q}}$, a large number of experimental tests have been done on all kinds of data.
\subsubsection{Filtering the synthetic data}
\begin{figure}[htbp]
  \centering
   \subfigure[Results due to LoG and Gaussian filters]{
                            \centering
        \includegraphics[width=4cm]{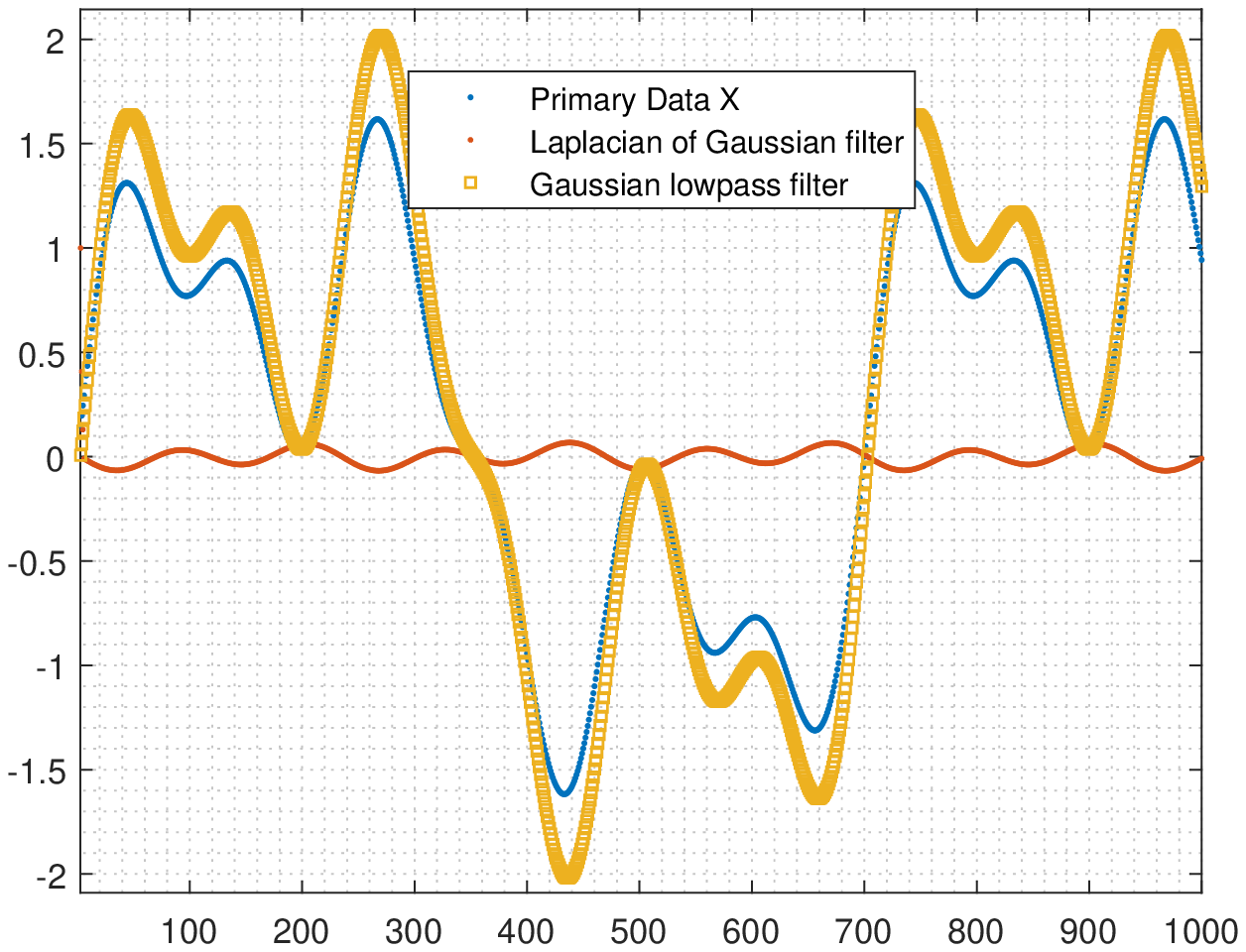}
}
   \subfigure[$\mathcal{I}_{m}(\mathcal{D}^{1+i}_{7}\mathbf{X}(t))$ and $\mathcal{M}_{o}(\mathcal{D}^{1+i}_{7}\mathbf{X}(t))$]{
                            \centering
        \includegraphics[width=4cm]{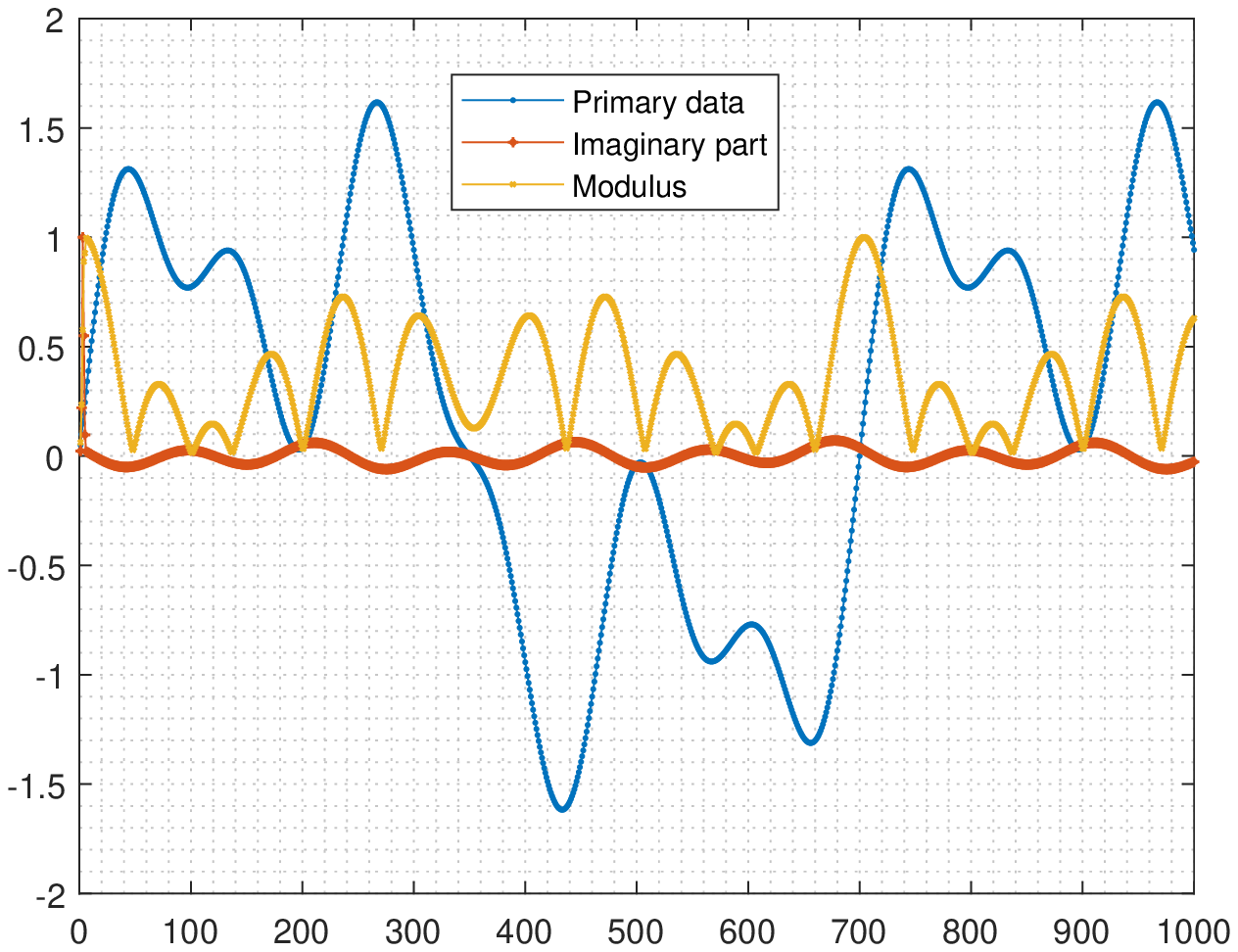}
}
   \subfigure[$\mathcal{A}_{n}(\mathcal{D}^{1+i}_{7}\mathbf{X}(t))$ and $\mathcal{R}_{e}(\mathcal{D}^{1+i}_{7}\mathbf{X}(t))$]{
                            \centering
        \includegraphics[width=4cm]{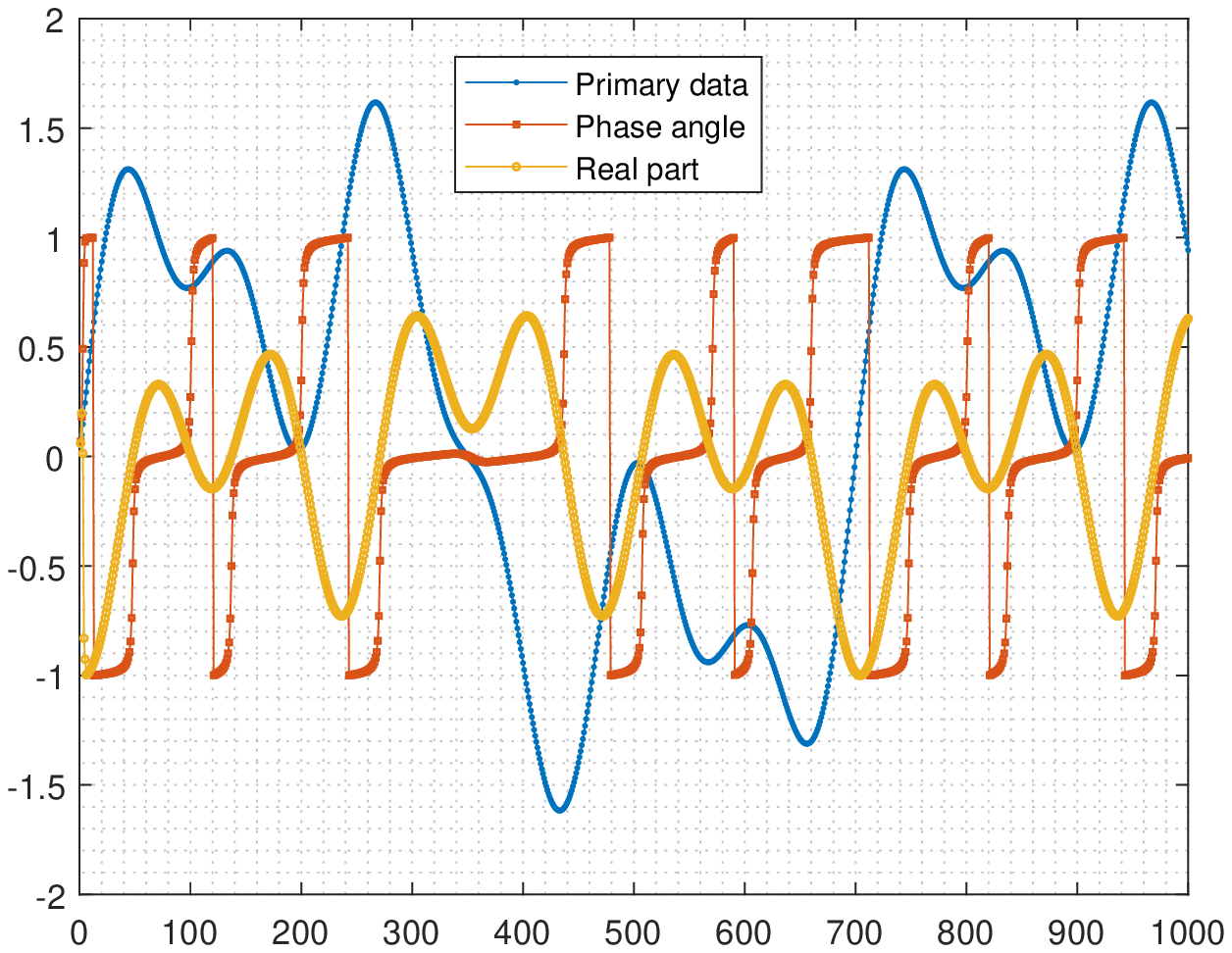}
}
\subfigure[$\mathcal{R}_{e}(\mathcal{I}^{1+i}_{7}\mathbf{X}(t))$ and $\mathcal{I}_{m}(\mathcal{I}^{1+i}_{7}\mathbf{X}(t))$]{
                            \centering
        \includegraphics[width=4cm]{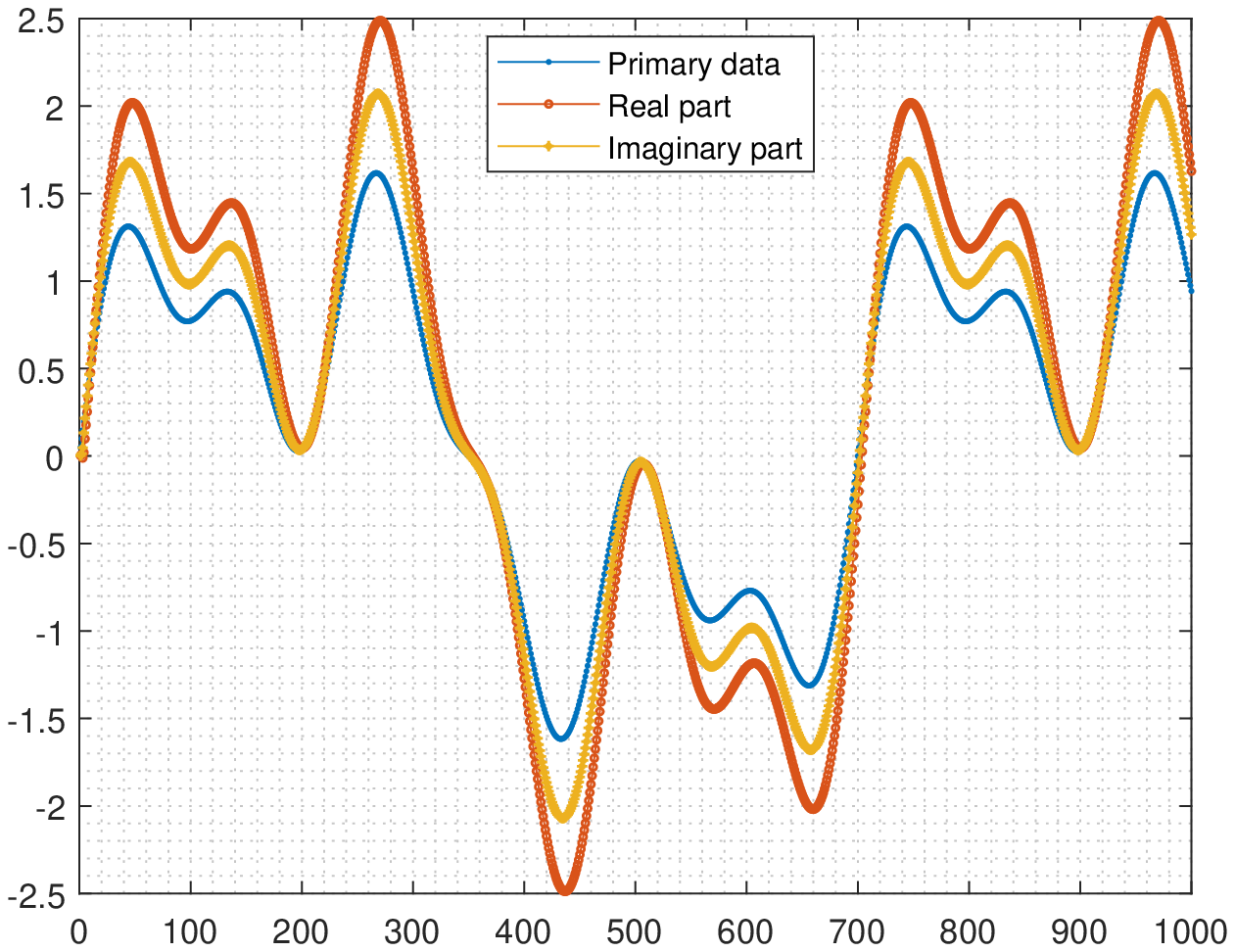}
}
  \caption{The synthetic data $\mathbf{X}(t)$ as well as its filtered results, $\mathcal{D}^{1+i}_{7}\mathbf{X}(t)$ and $\mathcal{I}^{1+i}_{7}\mathbf{X}(t)$ comparing the results due to the known filters, LoG and Gaussian.}\label{Fig.synthetic.data}
\end{figure}

In order to demonstrate the capability of $\mathcal{D}^{\alpha}_{n}$ to check changes with different frequencies, as well as the smooth capability of $\mathcal{I}^{\alpha}_{n}$, we construct a synthetic data sequence as follows
\bea \nonumber
\mathbf{X}(t)=\sin(\frac{2\pi t}{700})+0.7\sin(\frac{6\pi t}{700})+0.4\sin(\frac{12\pi t}{700}),t=1,\ldots10^3
\eea
which has three frequencies, and the components with higher frequency number have less amplitudes. We let $\alpha=1+i$ and $n=7$ without any special aim, only for showing the relations between the primary data sequence and the components of the filtered result, $\mathcal{D}_{7}^{1+i}$ and $\mathcal{I}_{7}^{1+i}$, as shown in Figure \ref{Fig.synthetic.data}.

In Figure \ref{Fig.synthetic.data}, comparing sub-figure (a) to (b) tells that, the imaginary part of $\mathcal{D}^{1+i}_{7}\mathbf{X}(t)$, $\mathcal{I}_{m}(\mathcal{D}^{1+i}_{7}\mathbf{X}(t))$, looks quite similar to the result of the filter, LoG. Of course, the results other than the imaginary part, shown in sub-figure (b) and (c), are information not provided by LoG. The congruent relationships between $\mathbf{X}(t)$ and $\mathcal{D}^{1+i}_{7}\mathbf{X}(t)$ seem complex: 1)$\mathcal{R}_{e}(\mathcal{D}^{1+i}_{7}\mathbf{X}(t))$ gets to the extremes about at the inflection points of $\mathbf{X}(t)$, and 2) the phase angle of $\mathcal{D}^{1+i}_{7}\mathbf{X}(t)$ has big gap values at the inflection points where $\mathbf{X}(t)$ varies from concave to convex. Hence, integrating sub-figure (b) and (c) can tells more details of $\mathbf{X}(t)$ than the result due to LoG filter, which is only a fraction of $\mathcal{D}^{1+i}_{7}\mathbf{X}(t)$. Comparing sub-figure (d) with (a), we can see $\mathcal{R}_{e}(\mathcal{I}^{1+i}_{7}\mathbf{X}(t))$ and $\mathcal{I}_{m}(\mathcal{I}^{1+i}_{7}\mathbf{X}(t))$ seem alike to the result due to the filter, Gaussian. In all, $\mathcal{D}^{1+i}_{7}\mathbf{X}(t)$ and $\mathcal{I}^{1+i}_{7}\mathbf{X}(t)$ uncover more secrets covered by $\mathbf{X}(t)$ than traditional filters designed in real domain, such as LoG and Gaussian filters.

\subsubsection{Filtering the benchmark visible-light image}
\begin{figure}[htbp]
  \centering
   \subfigure[Lena image $\mathbf{X}_{Lena}$]{
                            \centering
        \includegraphics[width=4cm]{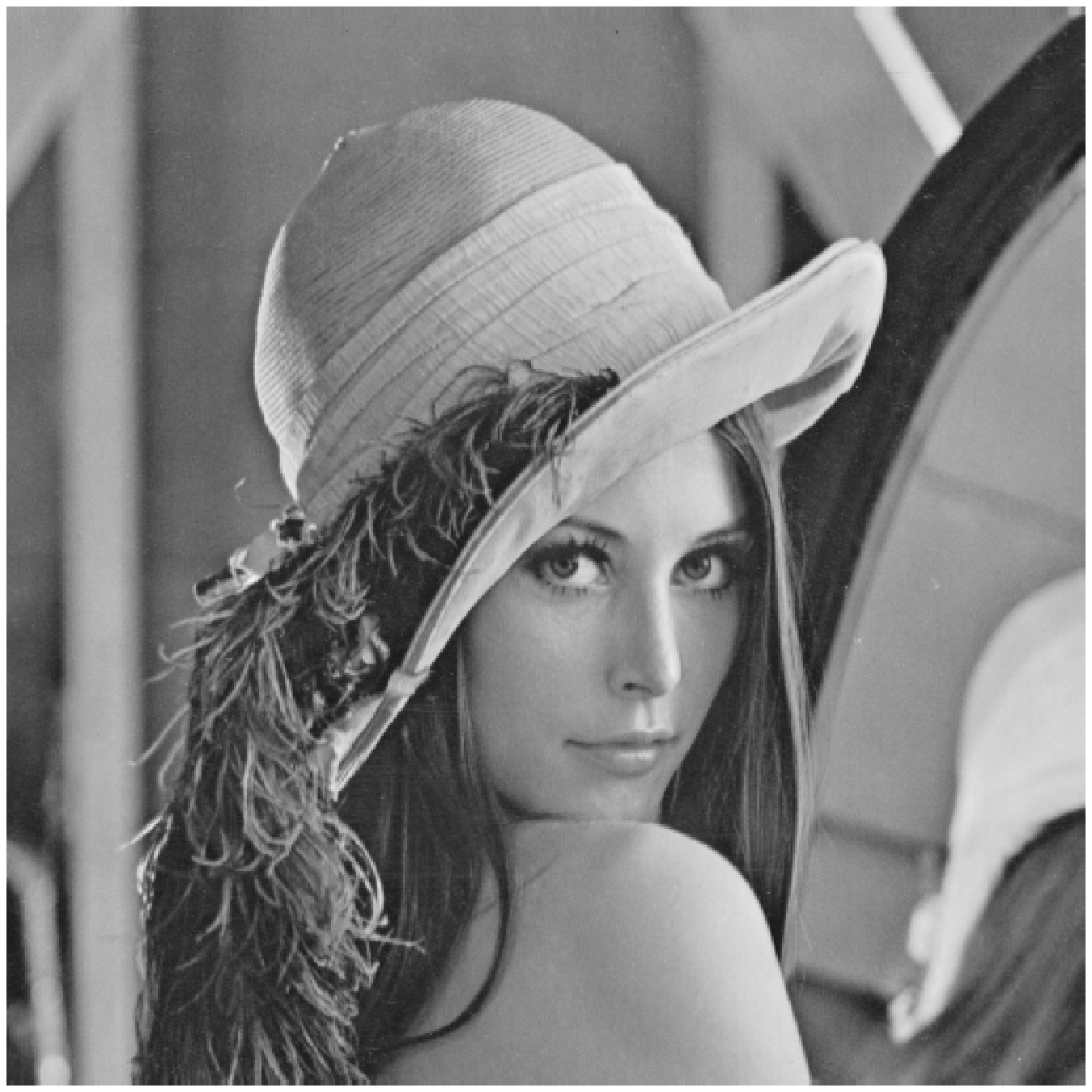}
}
   \subfigure[LoG of $\mathbf{X}_{Lena}$]{
                            \centering
        \includegraphics[width=4cm]{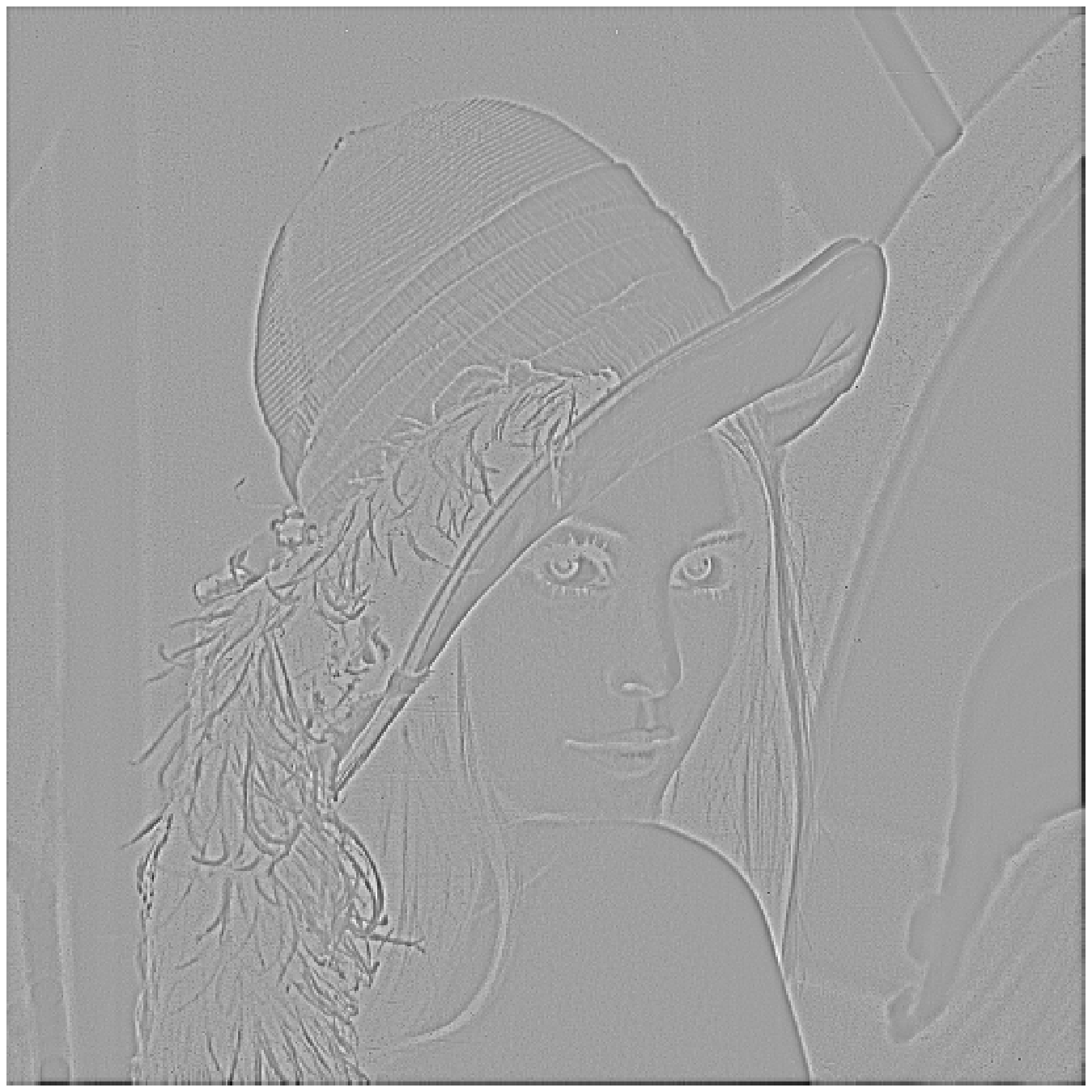}
}
  \caption{The primary Lena image $\mathbf{X}_{Lena}$ as well as the filtered result due to LoG}\label{Fig.lenaimage.LOG}
\end{figure}
Intensity images are 2-Dimensional signals, so without any lost of generalization, we use $\mathcal{D}^{1+i}_{7\times 7}$ and $\mathcal{I}^{1+i}_{7\times 7}$ to filter the image `Lena'\footnote{https://www.ece.rice.edu/~wakin/images/}, denoted as $\mathbf{X}_{Lena}$, which is very popular in image processing filed, somewhere of which has high frequency while someplace has low frequency.

\begin{figure}[htbp]
  \centering
   \subfigure[$\mathcal{R}_{e}(\mathcal{D}^{1+i}_{7\times7}\mathbf{X}_{Lena})$]{
                            \centering
        \includegraphics[width=4cm]{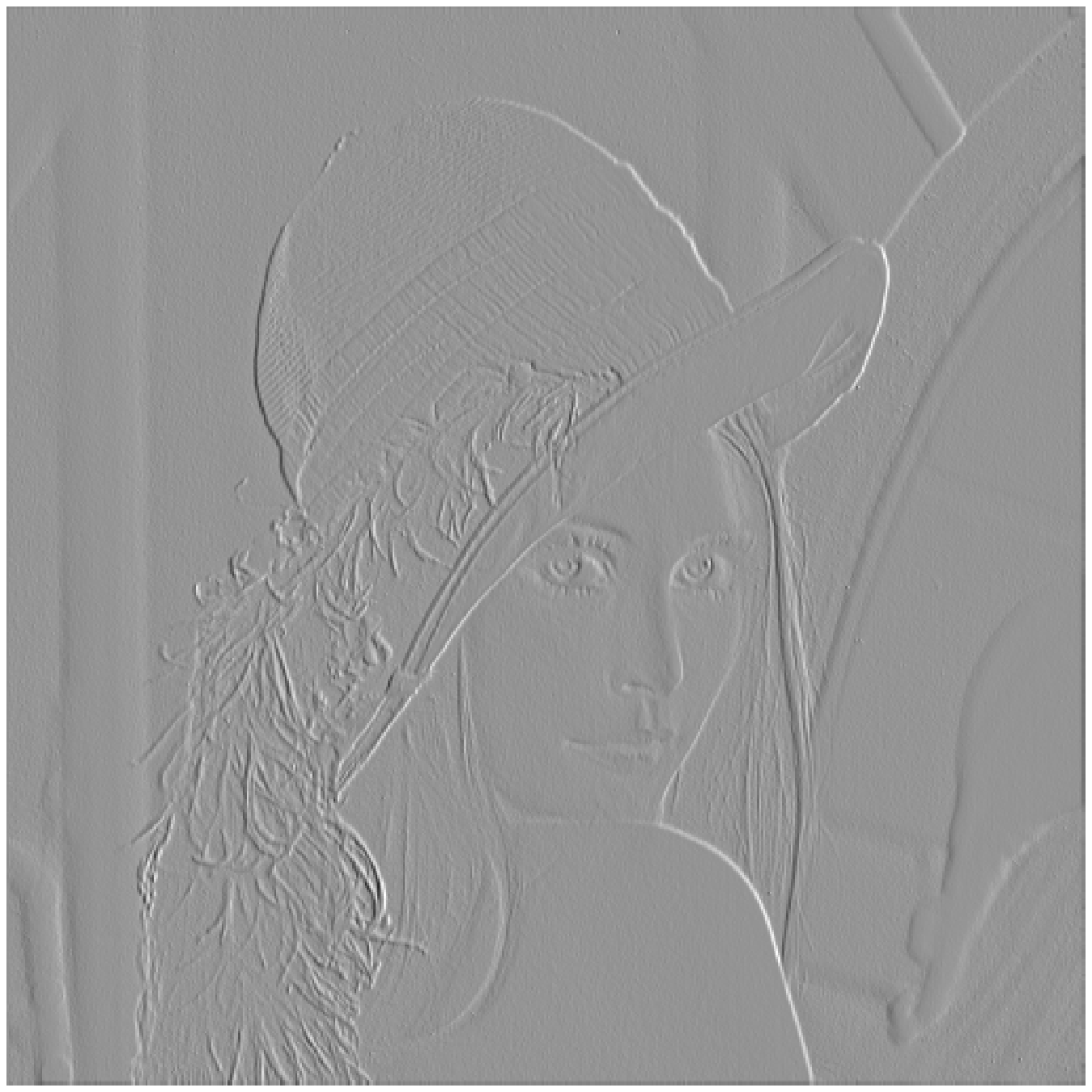}
}
   \subfigure[$\mathcal{I}_{m}(\mathcal{D}^{1+i}_{7\times7}\mathbf{X}_{Lena})$]{
                            \centering
        \includegraphics[width=4cm]{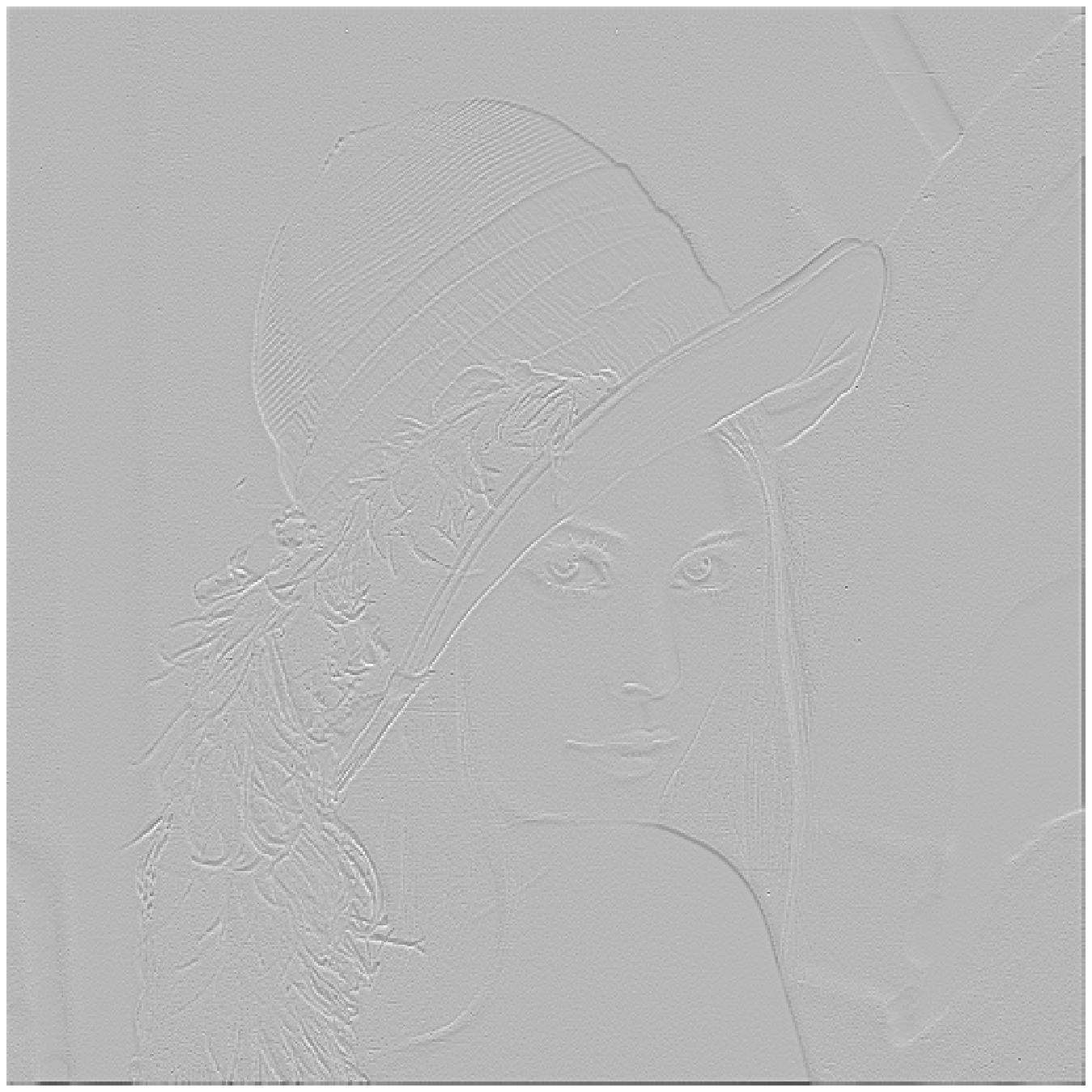}
}
   \subfigure[$\mathcal{A}_{n}(\mathcal{D}^{1+i}_{7\times7}\mathbf{X}_{Lena})$]{
                            \centering
        \includegraphics[width=4cm]{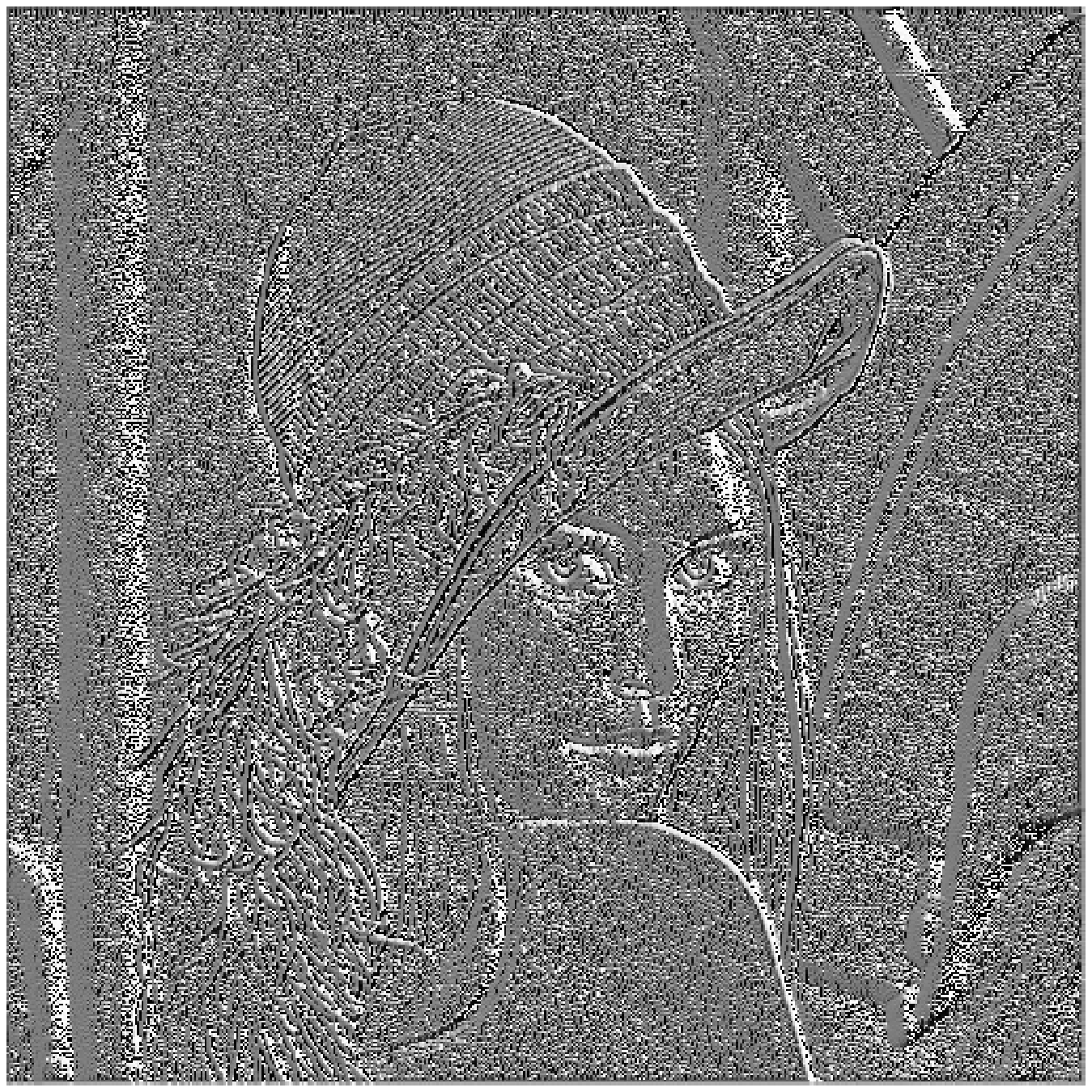}
}
\subfigure[The complementary image of $\mathcal{M}_{o}(\mathcal{D}^{1+i}_{7\times7}\mathbf{X}_{Lena})$]{
                            \centering
        \includegraphics[width=4cm]{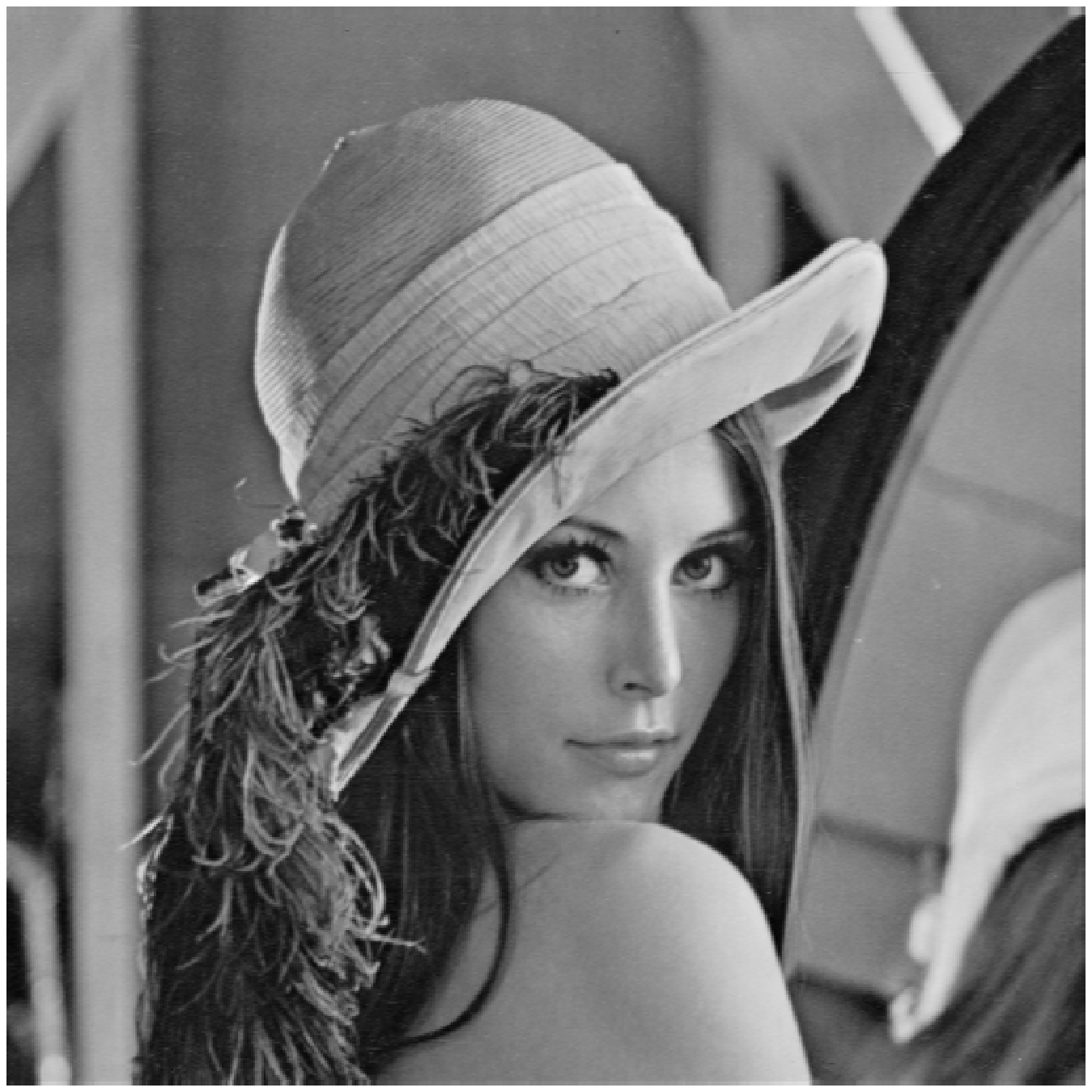}
}
  \caption{Components of $\mathcal{D}^{1+i}_{7\times7}\mathbf{X}_{Lena}$}\label{Fig.lenaimage.D}
\end{figure}

\begin{figure}[htbp]
  \centering
   \subfigure[$\mathcal{A}_{n}(\mathcal{I}^{1+i}_{7\times7}\mathbf{X}_{Lena})$]{
                            \centering
        \includegraphics[width=4cm]{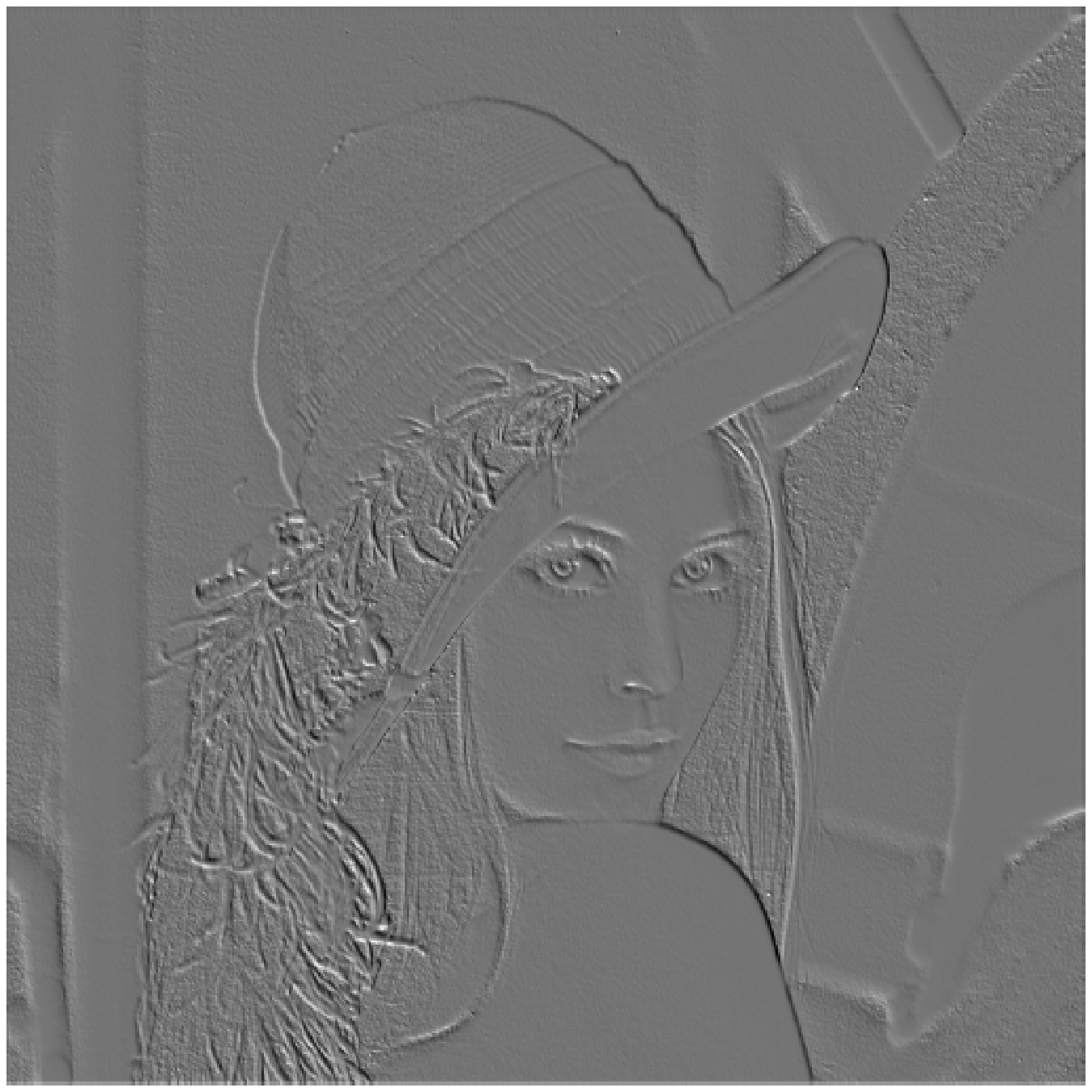}
}
   \subfigure[$\mathcal{M}_{o}(\mathcal{I}^{1+i}_{7\times7}\mathbf{X}_{Lena})$]{
                            \centering
        \includegraphics[width=4cm]{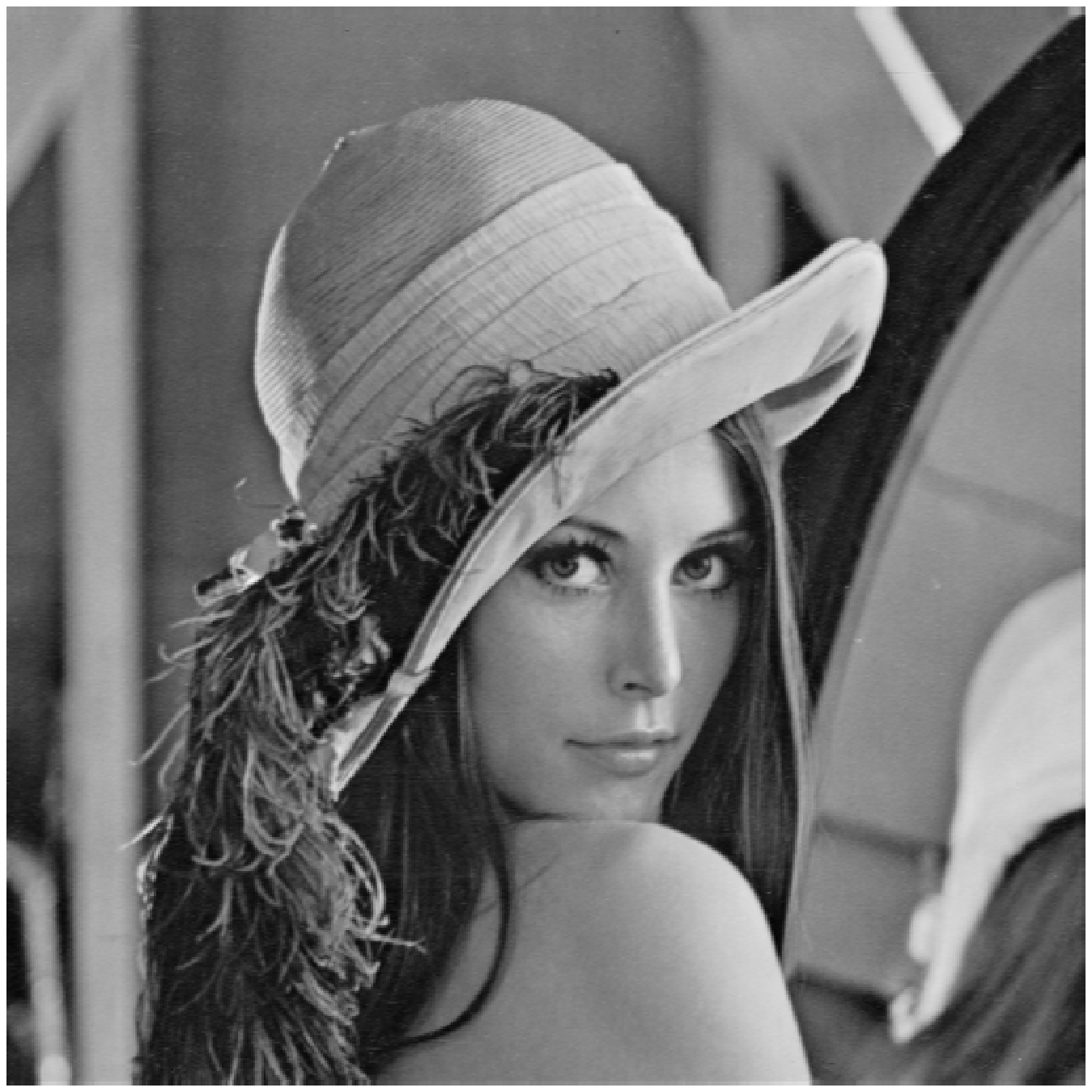}
}
  \caption{some components of $\mathcal{I}^{1+i}_{7\times7}\mathbf{X}_{Lena}$, the others, $\mathcal{R}_{e}(\mathcal{I}^{1+i}_{7\times7}\mathbf{X}_{Lena})$ and $\mathcal{I}_{m}(\mathcal{I}^{1+i}_{7\times7}\mathbf{X}_{Lena})$ seem very like to the primary data, as shown in subfigure (a) of Figure \ref{Fig.lenaimage.LOG}}\label{Fig.lenaimage.I}
\end{figure}

First, we use a LoG filter with standard deviation $0.5$ and size $7\times 7$ to filter $\mathbf{X}_{Lena}$, the result is shown in Sub-figure (b) of Figure \ref{Fig.lenaimage.LOG}. The Filtered results of $\mathcal{D}^{1+i}_{7\times 7}$ and $\mathcal{I}^{1+i}_{7\times 7}$ are shown in Figure \ref{Fig.lenaimage.D} and \ref{Fig.lenaimage.I}. Comparing Sub-figure (b) of Figure \ref{Fig.lenaimage.LOG} with Figure \ref{Fig.lenaimage.D}, we can see that: 1) $\mathcal{R}_{e}(\mathcal{D}^{1+i}_{7\times7}\mathbf{X}_{Lena})$ and $\mathcal{I}_{m}(\mathcal{D}^{1+i}_{7\times7}\mathbf{X}_{Lena})$ and sub-figure (b) of Figure \ref{Fig.lenaimage.LOG} are very like with each other, and the complementary image of $\mathcal{M}_{o}(\mathcal{D}^{1+i}_{7\times7}\mathbf{X}_{Lena})$ (sub-figure (d) in Figure \ref{Fig.lenaimage.D}) shows the sketch of Lena; 2) $\mathcal{A}_{n}(\mathcal{D}^{1+i}_{7\times7}\mathbf{X}_{Lena})$, as shown in sub-figure (c) in Figure \ref{Fig.lenaimage.D}, has so many noises, and this result seems to demonstrate that $\mathcal{A}_{n}(\mathcal{D}^{1+i}_{7\times7}\mathbf{X}_{Lena})$ is very sensitive to the changes in $\mathbf{X}_{Lena}$, even in the relative smooth areas there are so many noise points, and this phenomenon is consistent with that, phase correlation is very sensitive to noises. Comparing sub-figure (a), (b) and (c) with each other we can see they are different, and a bit complementary with each other. Comparing Figure \ref{Fig.lenaimage.I} with Sub-figure (a) of Figure \ref{Fig.lenaimage.LOG}, we can see $\mathcal{A}_{n}(\mathcal{I}^{1+i}_{7\times7}\mathbf{X}_{Lena})$ seems to enhance the area where relative larger changes happen; $\mathcal{M}_{o}(\mathcal{I}^{1+i}_{7\times7}\mathbf{X}_{Lena})$ seems very like to $\mathbf{X}_{Lena}$, and too are $\mathcal{R}_{e}(\mathcal{I}^{1+i}_{7\times7}\mathbf{X}_{Lena})$ and $\mathcal{I}_{m}(\mathcal{I}^{1+i}_{7\times7}\mathbf{X}_{Lena})$ actually.

\subsubsection{Filtering TRUS (TRansrectal UltraSound) images}

\begin{figure}[htbp]
  \centering
   \subfigure[TRUS image $\mathbf{X_{1}}_{trus}$]{
                            \centering
        \includegraphics[width=4cm]{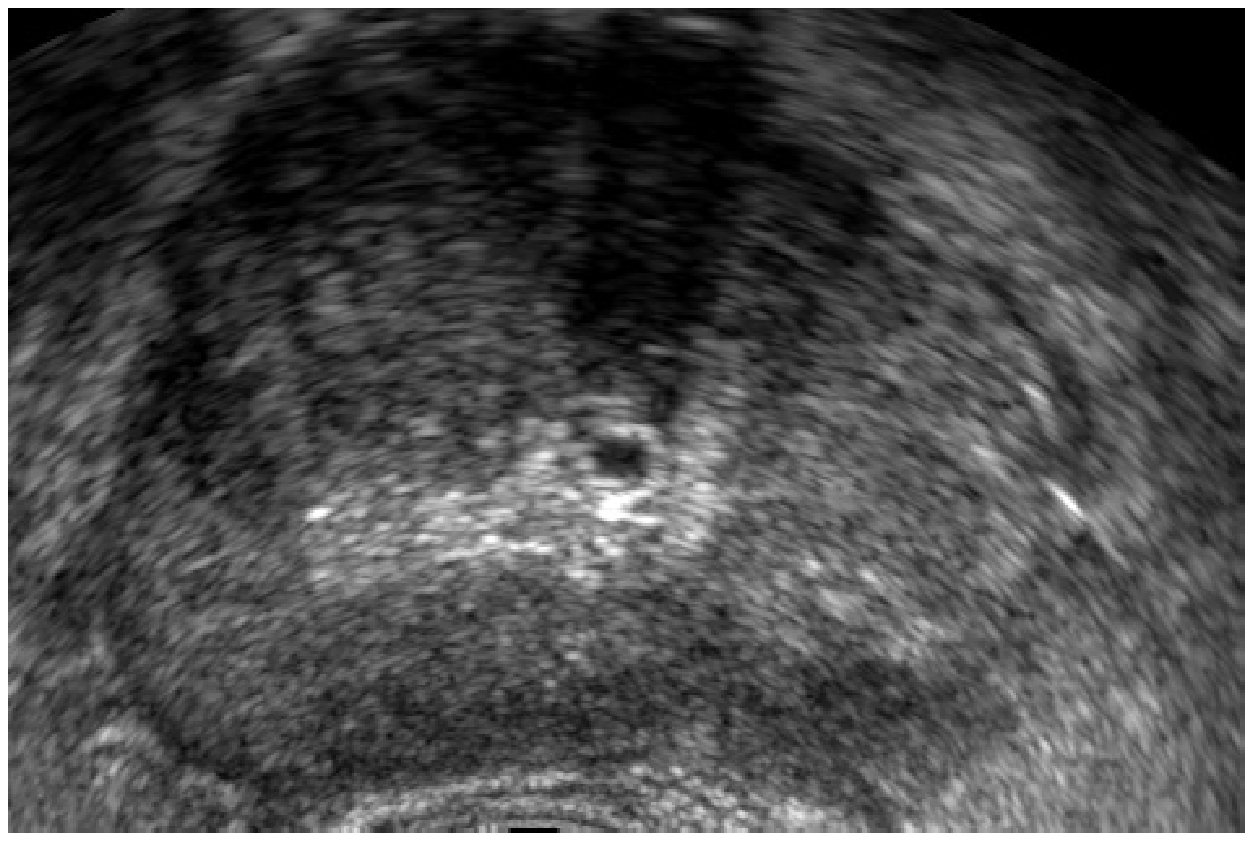}
}
   \subfigure[Gaussian filter of $\mathbf{X_{1}}_{trus}$]{
                            \centering
        \includegraphics[width=4cm]{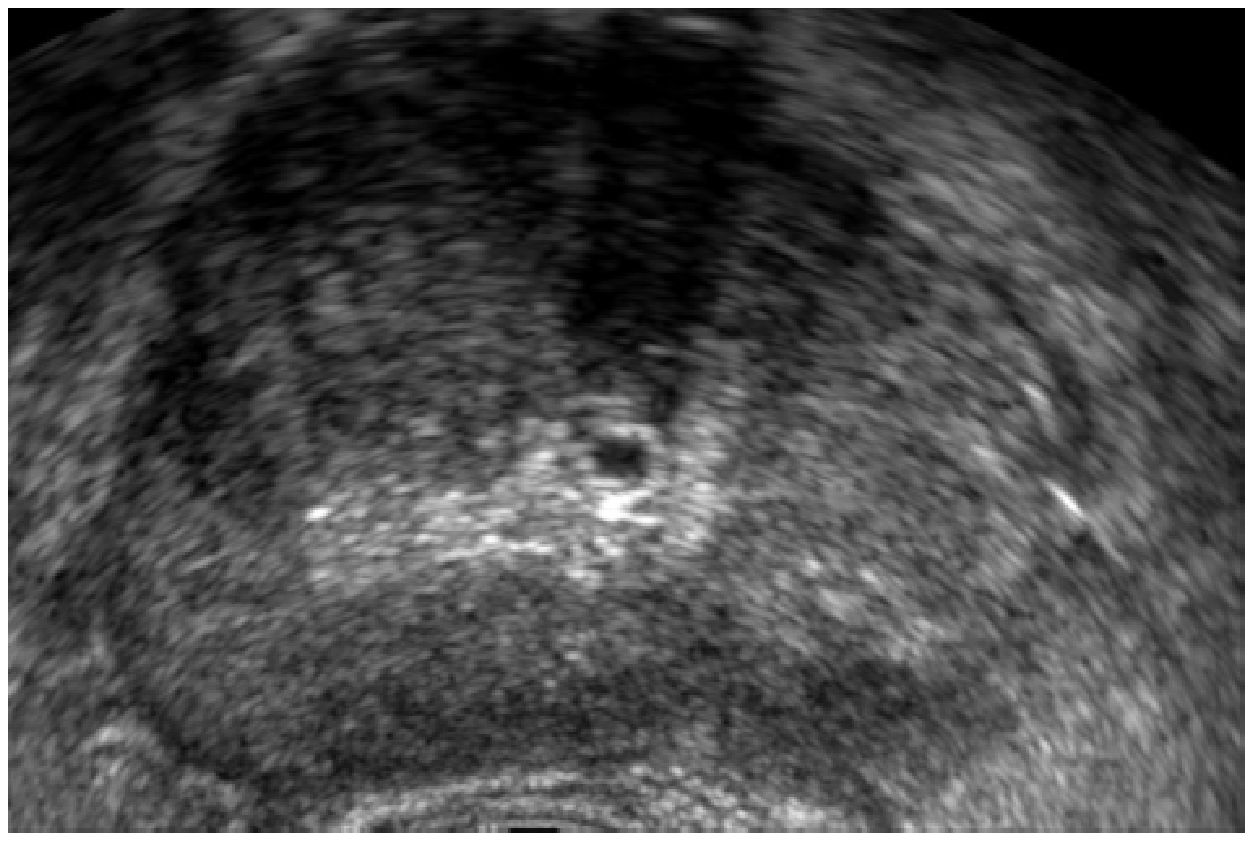}
}
  \caption{TRUS image, $\mathbf{X_{1}}_{trus}$,  has heavy speckles, and the Gaussian low-pass filter cannot reveal any useful cue}\label{Fig.TRUS1.prim.gaussian}
\end{figure}
\begin{figure}[htbp]
  \centering
   \subfigure[$\mathcal{R}_{e}(\mathcal{I}^{1+i}_{7\times7}\mathbf{X_{1}}_{trus})$]{
                            \centering
        \includegraphics[width=4cm]{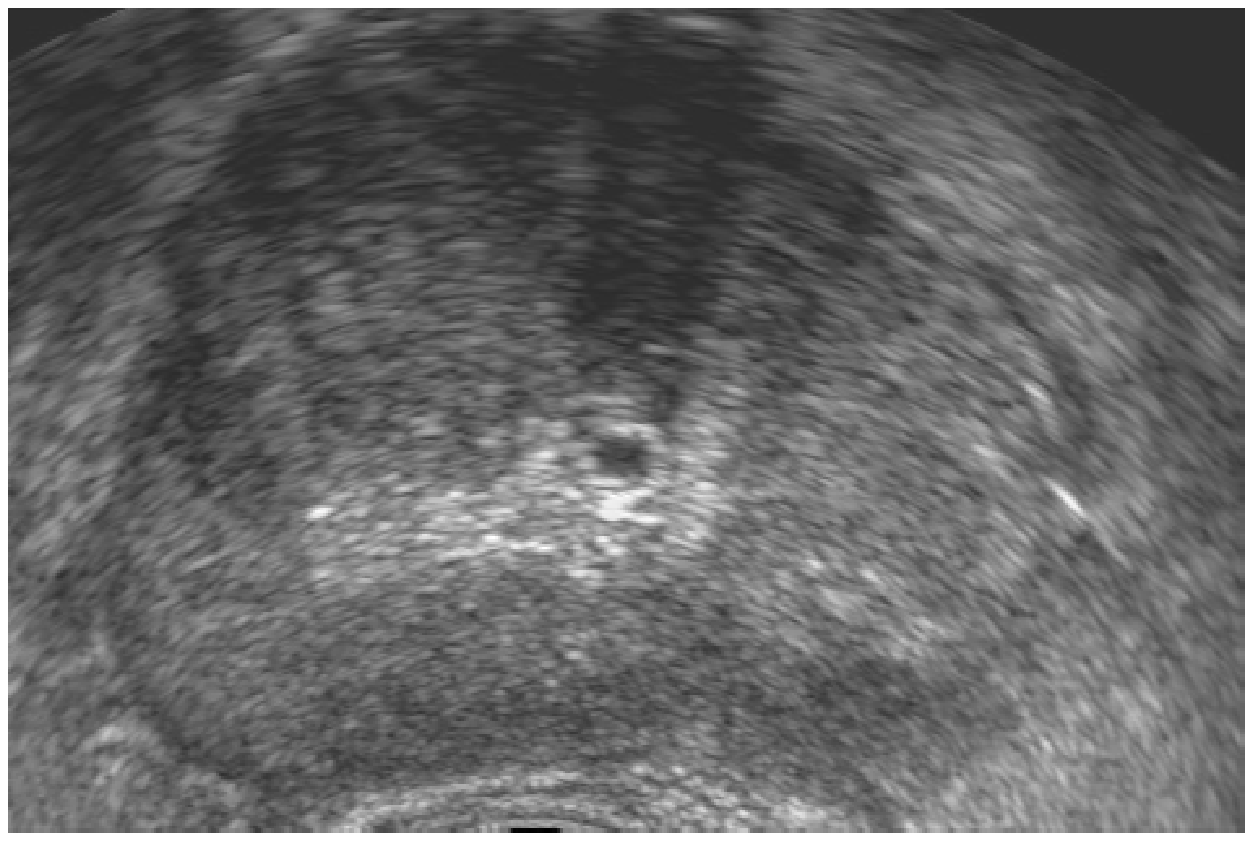}
}
   \subfigure[$\mathcal{I}_{m}(\mathcal{I}^{1+i}_{7\times7}\mathbf{X_{1}}_{trus})$]{
                            \centering
        \includegraphics[width=4cm]{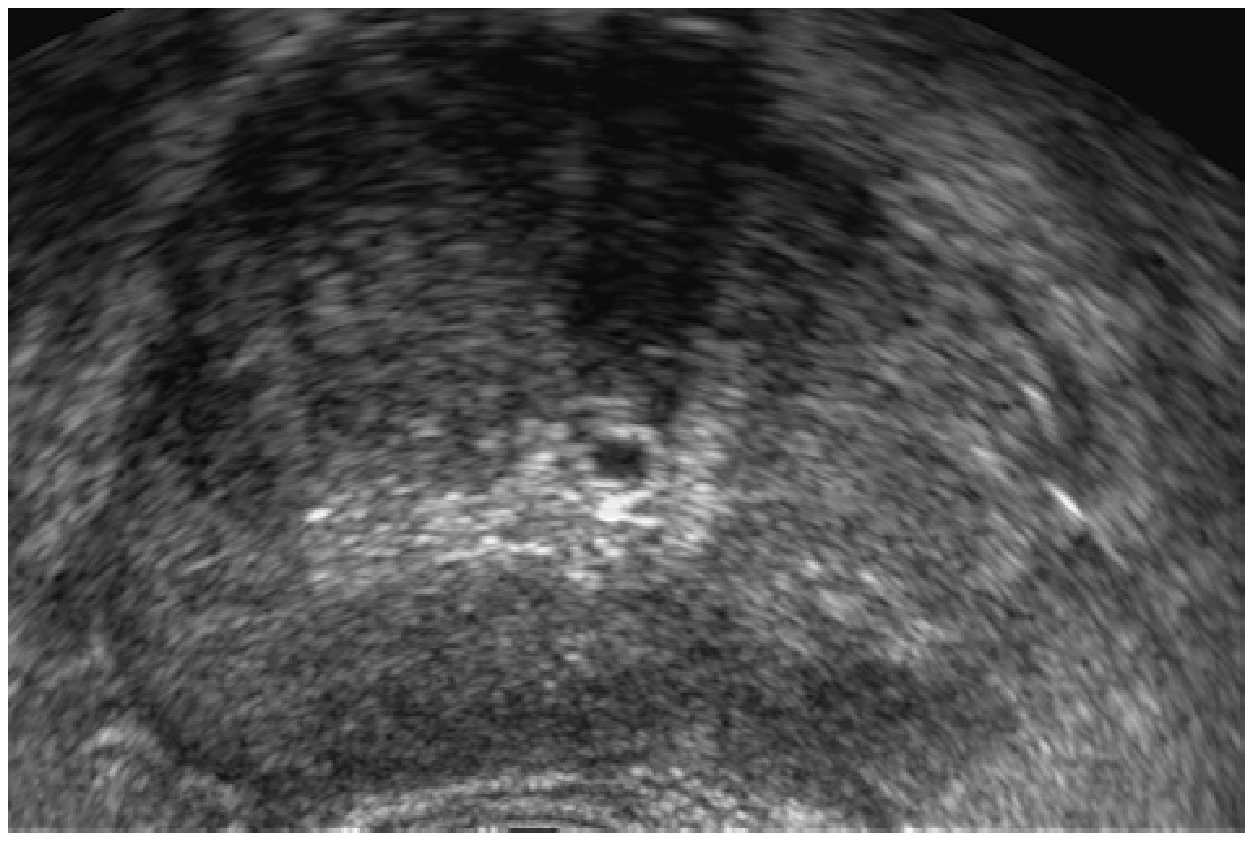}
}
   \subfigure[$\mathcal{A}_{n}(\mathcal{I}^{1+i}_{7\times7}\mathbf{X_{1}}_{trus})$]{
                            \centering
        \includegraphics[width=4cm]{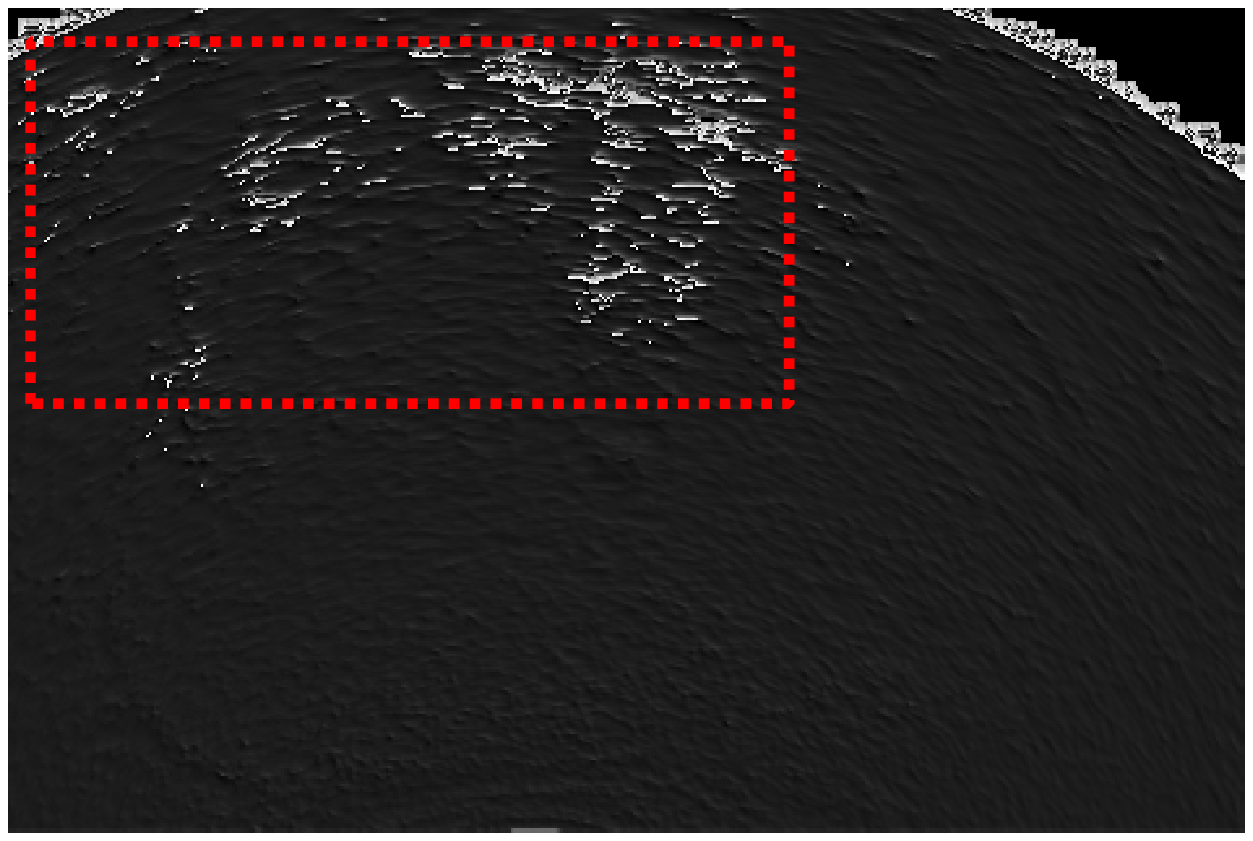}
}
\subfigure[$\mathcal{M}_{o}(\mathcal{I}^{1+i}_{7\times7}\mathbf{X_{1}}_{trus})$]{
                            \centering
        \includegraphics[width=4cm]{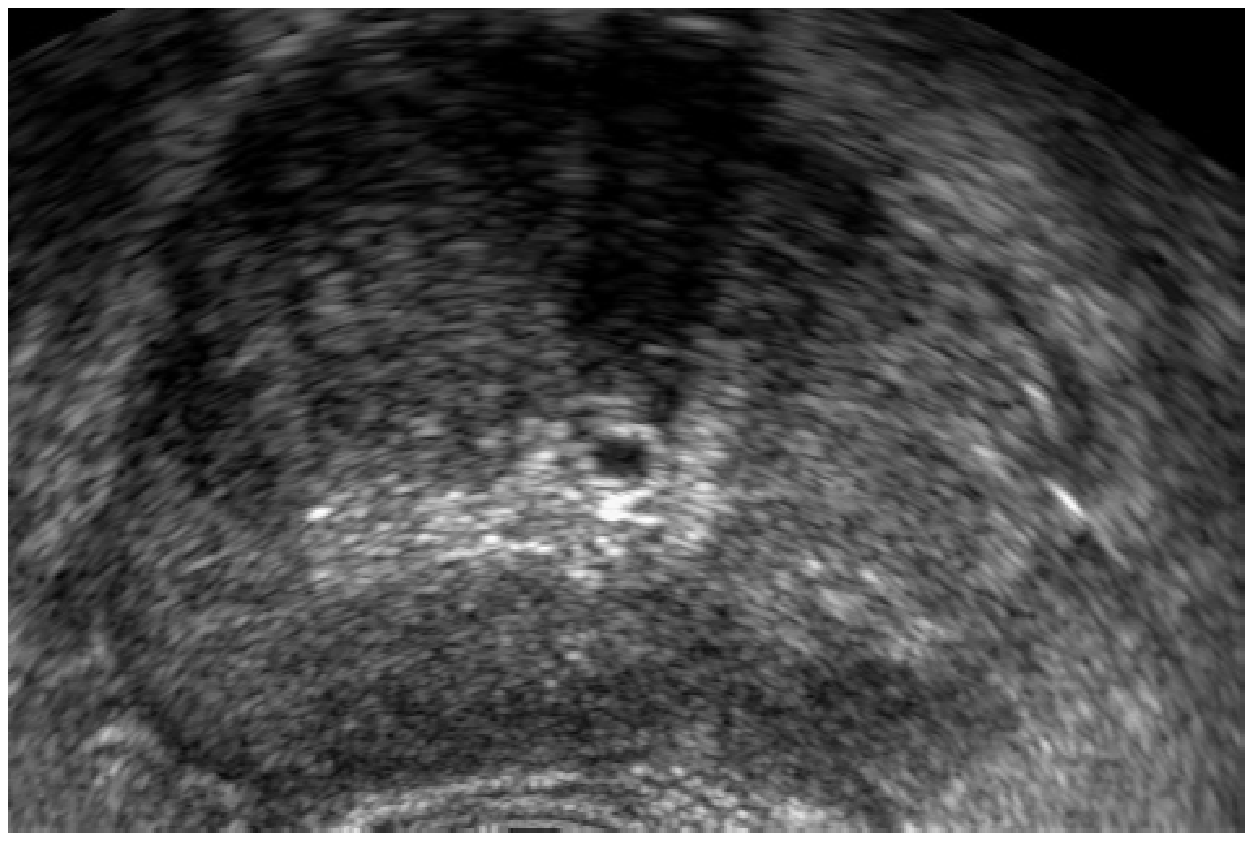}
}
  \caption{Components of $\mathcal{I}^{1+i}_{7\times7}\mathbf{X_{1}}_{trus}$, and $\mathcal{A}_{n}(\mathcal{I}^{1+i}_{7\times7}\mathbf{X_{1}}_{trus})$ seems amazing to show the skeptical region }\label{Fig.trusimage.I}
\end{figure}

TRUS image is the main means for prostate cancer diagnosis and treatment,but this is always hampered by heavy speckles, as shown in subfigure (a) of Figure \ref{Fig.TRUS1.prim.gaussian}. How to find the diseased region and segment the prostate is rather challenging. Because there is no effective way to automatically complete the task, and it is still mainly performed by manual work up to now\cite{1046}. Due to heavy speckle noise, usually gradient operations do not be valid in disclosing the prostate boundary or in finding the diseased positions, though they behave admissibly for visible-light images. Using smooth filters to depress the speckles possibly benefit the analysis of the images, however from sub-figure (b) of Figure \ref{Fig.TRUS1.prim.gaussian} as well as sub-figure (a), (b) and (d) of Figure \ref{Fig.trusimage.I}, we can see the smoothed results still have heavy speckles, and do not highlight any image features which make the diseased area and boundary prominent. So, Gaussian smoothing, $\mathcal{R}_{e}(\mathcal{I}^{1+i}_{7\times7}\mathbf{X_{1}}_{trus})$, $\mathcal{I}_{m}(\mathcal{I}^{1+i}_{7\times7}\mathbf{X_{1}}_{trus})$ as well as $\mathcal{M}_{o}(\mathcal{I}^{1+i}_{7\times7}\mathbf{X_{1}}_{trus})$ seem useless to identify the region of interest for TRUS images.

\begin{figure}[htbp]
  \centering
   \subfigure[TRUS image $\mathbf{X_{2}}_{trus}$]{
                            \centering
        \includegraphics[width=4cm]{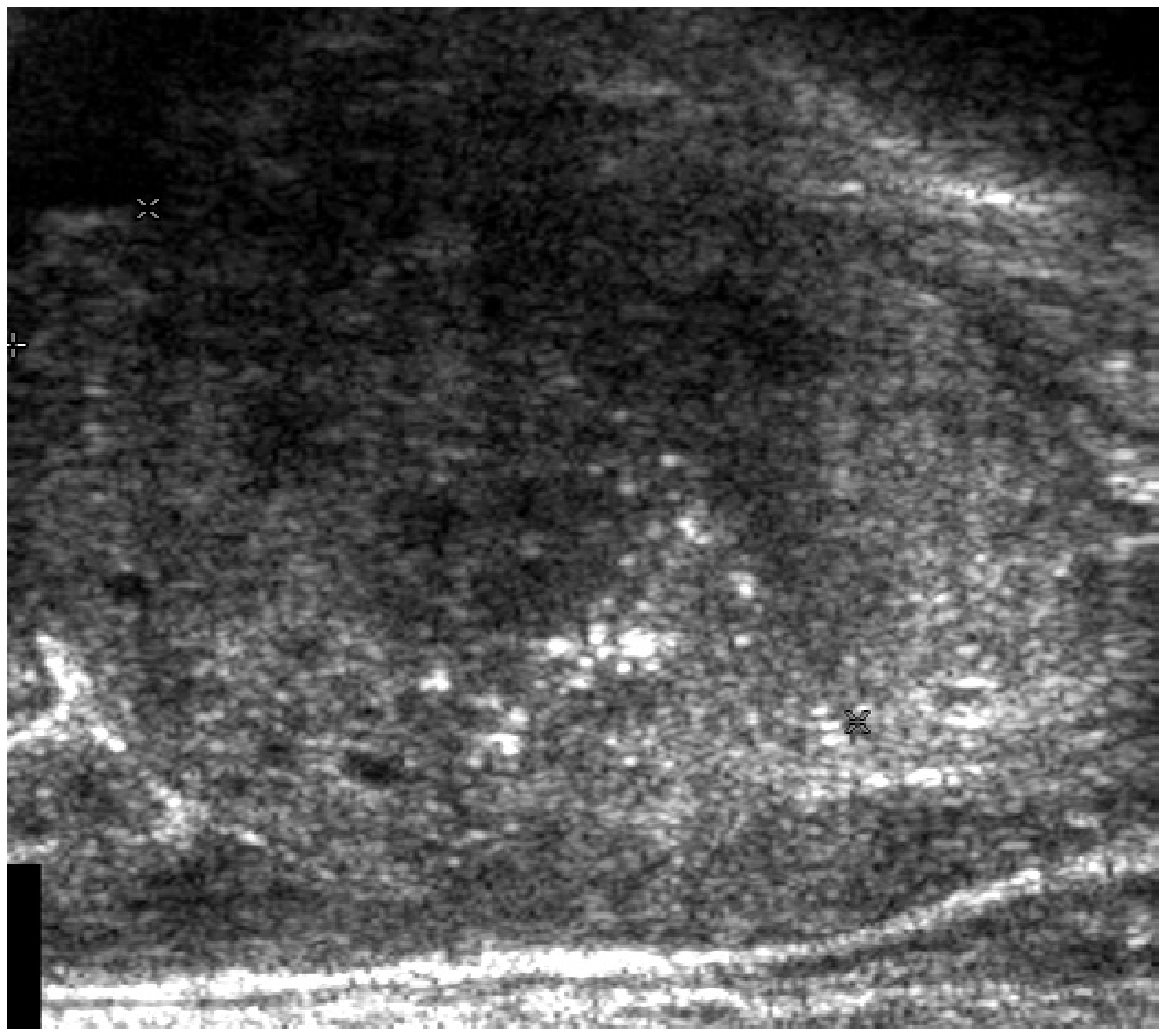}
}
   \subfigure[$\mathcal{A}_{n}(\mathcal{I}^{1+i}_{7\times7}\mathbf{X_{2}}_{trus})$]{
                            \centering
        \includegraphics[width=4cm]{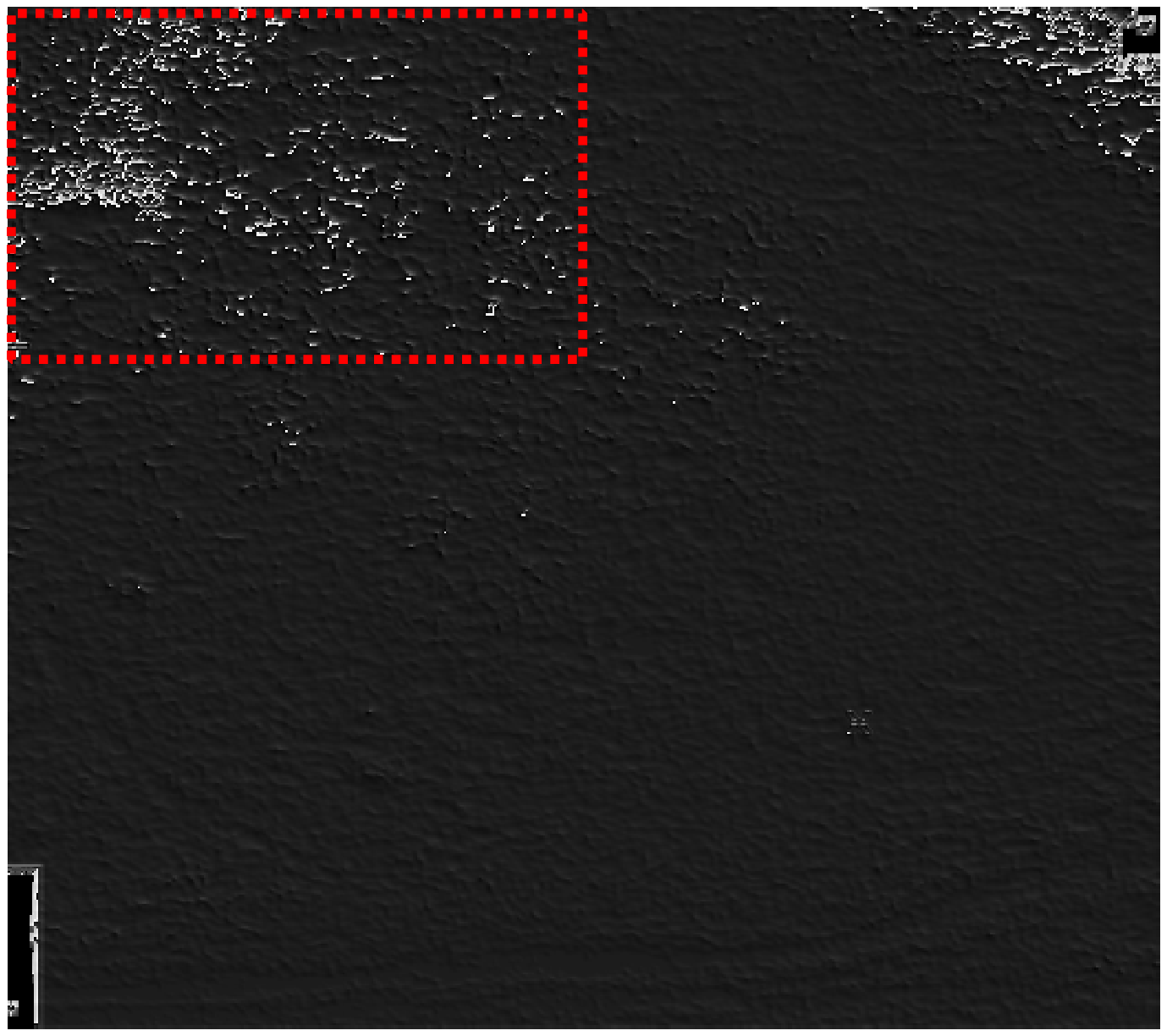}
}
  \caption{TRUS image, $\mathbf{X_{2}}_{trus}$ as well as its $\mathcal{A}_{n}(\mathcal{I}^{1+i}_{7\times7}\mathbf{X_{2}}_{trus})$}\label{Fig.TRUS2.prim.I}
\end{figure}

However, $\mathcal{A}_{n}(\mathcal{I}^{1+i}_{7\times7}\mathbf{X_{1}}_{trus})$, shown in sub-figure (c) of Figure  \ref{Fig.trusimage.I}, seems very beneficial for indicating the diseased or boundary region. From sub-figure (c) of Figure \ref{Fig.trusimage.I}, we can see that within the skeptical diseased regions there are so many extreme points of phase angles. Actually, in the region the prostate boundary is not clear, and it is in high probability that in the region tumour exists. Though $\mathcal{R}_{e}(\mathcal{I}^{1+i}_{7\times7}\mathbf{X_{1}}_{trus})$, $\mathcal{I}_{m}(\mathcal{I}^{1+i}_{7\times7}\mathbf{X_{1}}_{trus})$ cannot indicate this even a little, but $\mathcal{A}_{n}(\mathcal{I}^{1+i}_{7\times7}\mathbf{X_{1}}_{trus})$ obviously show this. It seems very amazing, using this $\mathcal{A}_{n}(\mathcal{I}^{1+i}_{7\times7}\mathbf{X_{1}}_{trus})$ may provide significant improvement in handling TRUS images. A large number of tests was additionally done, and the found phenomenon seems robust, for example, the result shown in Figure \ref{Fig.TRUS2.prim.I}, where sub-figure (b) also indicates the skeptical region. In all, complex-order filters seem very potential in handling images with heavy speckles.

\subsubsection{Discussions}
In all above tests, $\alpha$ takes $1+i$, and the results, especially the ones on TRUS images, behave great potentially and promisingly. Of course, how to evaluate $\alpha$ is vital for a real application. This problem is important, but up to now, there is no way to determine the appropriate calculus order $\alpha$ for a real problem. So this problem is not discussed here. When $\alpha$ takes complex number, $\mathcal{D}^{\alpha}_{n_{1}\times n_{2}\ldots\times n_{q}}\mathbf{X}$ and $\mathcal{I}^{\alpha}_{n_{1}\times n_{2}\ldots\times n_{q}}\mathbf{X}$ usually are complex numbers. Compared with the real $\alpha$, which makes $\mathcal{D}^{\alpha}_{n_{1}\times n_{2}\ldots\times n_{q}}\mathbf{X}$ and $\mathcal{I}^{\alpha}_{n_{1}\times n_{2}\ldots\times n_{q}}\mathbf{X}$ to be real numbers according to Theorem \ref{theorem2}, complex $\alpha$ makes  $\mathcal{D}^{\alpha}_{n_{1}\times n_{2}\ldots\times n_{q}}\mathbf{X}$ and $\mathcal{I}^{\alpha}_{n_{1}\times n_{2}\ldots\times n_{q}}\mathbf{X}$ be complex numbers, and the real and imaginary parts, phase angle and modulus can be used to study $\mathbf{X}$. In some scenarios, phase angle behave amazingly compared with real and imaginary parts, as shown in sub-figure (b) of Figure \ref{Fig.TRUS2.prim.I} and sub-figure (c) of Figure \ref{Fig.lenaimage.I}. We can think, appropriately fusing real and imaginary parts may produce more meaningful apparatuses in practice.
\section{Conclusions} \label{sec.con}
In this paper, the filters with complex derivative or integral order are proposed. By multi-section trick, the complex-order filter has been proved to be the superposition of segments, gotten by partitioning the expansion sequence of the complex-order binomial. Compared with real number order filters, the complex order ones can uncover more information hidden in data sequence. A large number of tests  have shown that the filters behave potentially and promisingly in disclosing features of data. Especially for the challenging problem of transrectal ultrasound images, the complex-order filters can indicate the diseased regions, and this cannot be done usually for real-order filters. In all, it is very hopeful that the complex-order filters will significantly benefit image processing and the associated fields
\section*{Acknowledgment}
This work is supported by NSFC under grants 61860206007 and U19A2071.
\bibliographystyle{ieeetr}
\bibliography{ref}
\end{document}